\documentclass[%
 reprint,
superscriptaddress,
amsmath,amssymb,
aps,
prb,
longbibliography,nofootinbib
]{revtex4-2}

\usepackage{graphicx}
\usepackage{dcolumn}
\usepackage{bm,bbm,mathrsfs}
\usepackage{soul}
\usepackage{xcolor}
\usepackage[colorlinks=true,linkcolor=blue,citecolor=blue]{hyperref}

\newcommand{\ket}[1]{\left|#1\right\rangle}

\newcommand{\inp}[2]{\langle#1|#2\rangle}

\begin{document}

\preprint{APS/123-QED}

\title{Scattering theory of delicate topological insulators}

 \author{Penghao Zhu} \affiliation{Department of Physics and Institute for Condensed Matter Theory, University of Illinois at Urbana-Champaign, Urbana, Illinois 61801, USA}
 \author{Jiho Noh}
 \affiliation{Department of Mechanical Science and Engineering,
University of Illinois at Urbana–Champaign, Urbana, IL 61801 USA}
\author{Yingkai Liu}
\affiliation{Department of Physics and Institute for Condensed Matter Theory, University of Illinois at Urbana-Champaign, Urbana, Illinois 61801, USA}
 \author{Taylor L. Hughes}
 \affiliation{Department of Physics and Institute for Condensed Matter Theory, University of Illinois at Urbana-Champaign, Urbana, Illinois 61801, USA}




\date{\today}

\begin{abstract}
We study the scattering theory of delicate topological insulators (TIs), which are novel topological phases beyond the paradigm of the tenfold way, topological quantum chemistry, and the symmetry indicator method. We demonstrate that the phase of the reflection amplitude can probe the delicate topology by capturing a characteristic feature of a delicate TI. This feature is the returning Thouless pump, where an integer number of charges are pumped forward and backward in the first and second half of the adiabatic cycle respectively. As a byproduct of our analysis we show that requiring additional symmetries can stable the boundary states of a delicate TI beyond the conventional requirement of a sharply defined surface. Furthermore, we propose a photonic crystal experiment to implement a delicate TI and measure its reflection phase which reveals the delicate topology. 
\end{abstract}
\maketitle

\section{Introduction}
Delicate topology has been introduced recently as a fine-grained classification of phases~\cite{nelson:2021multi,nelson:2021delicate} that were previously considered trivial in the tenfold way~\cite{kitaev2009periodic,chiu2016}, topological quantum chemistry~\cite{bradlyn2017topological,cano2018building,cano2021band}, and symmetry indicator~\cite{Fu2007,Hughes2011,Fu2011,Fang2012,Fang2013,slager2013space,Teo2013,Wladimir2014,Kruthoff2017,Miert2018,Wladimir2019,Li2020} classification schemes. The characteristic feature of a class of delicate topological insulators (TIs) is a $2\pi$-quantized difference between two Berry phases defined over a pair of high symmetry lines in the Brillouin zone (BZ), and it is delicate in the sense that this quantization can be nullified by adding trivial bands to either the occupied or unoccupied subspace. 

As we review below, since nontrivial Chern numbers are precluded in a delicate TI by assumption, the $2\pi$-quantized difference in the Berry phase does not indicate an adiabatic pumping of charge~\cite{thouless1983}, but instead generates a returning Thouless pump (RTP). Indeed,  if we regard the momentum in the direction perpendicular to the high symmetry lines as an adiabatic parameter, an RTP indicates that an integer number of charges are pumped toward one direction by an integer number of unit cells in the first half of the cycle, and are pumped back to their starting point in the second half of the cycle~\cite{nelson:2021multi,nelson:2021delicate,zhu2022}. A non-vanishing RTP in a delicate TI guarantees that gapless surface states will be localized on a \emph{sharp} boundary with no deformations/reconstructions, which is weaker than the conventional TI bulk-boundary correspondence. Such a sharp boundary is perhaps difficult to achieve naturally in solid-state materials, but could be engineered in metamaterials such as photonic or acoustic crystals. 

Besides the gapless surface states on a sharp boundary and previous work by some of us on surface magnetism \cite{zhu2021}, few robust experimental observables of delicate TIs have been proposed. Here we plan to leverage the scattering theory of topological phases to provide a practical route to extract delicate topological data from a reflection matrix \cite{fulga:2011scattering,fulga:2012scattering,Hu2015,Wang2017}. 
Previous studies have comprehensively discussed the scattering theory for stable topological phases protected by internal symmetries, i.e., phases in the tenfold way periodic table\cite{fulga:2012scattering}. Recently, the study of scattering theory has been extended to a wider range of topological phases, e.g., 2D higher-order TIs that have gapped edges but gapless corners \cite{selma:2021simulating,franca:2022topological}. However, the scattering theory of delicate TIs, which may provide new insight and experimental observables for delicate topology, is still absent. To address this, here we study the scattering theory of delicate TIs protected by rotation/mirror symmetry and propose a photonic experiment to measure our predicted reflection-phase observable.

The remainder of the article is organized as follows. In Sec. \ref{sec:review}, we give a brief review of delicate TIs. Then, in Sec. \ref{sec:pump}, we investigate the relationship between the number of pumped charges during an adiabatic process and the reflection phase. We show that the pumped charges that lead to an RTP in a delicate TI can be detected by the phase of the reflection amplitude on a sharp boundary. We also show,  for the first time, that if there are extra symmetries on the high symmetry lines in the BZ, then the boundary modes, and the nontrivial behavior of the reflection phase, can remain robust even when the requirement of a sharp boundary is relaxed.  Finally, in Sec. \ref{sec:expproposal} we propose a photonic experiment to implement a delicate TI in a synthetic dimension where measurements of the reflection phase are possible and can be used to detect the bulk RTP and hence delicate topology. We end our article with conclusions and remarks about open questions for future research in Sec. \ref{sec:conclusion}.

\section{Review of delicate topology, RTP, and surface states}
\label{sec:review}
Before discussing the scattering theory, we shall first review delicate topology and its characteristic RTP protected by rotation/mirror symmetry. For clarity, let us first set up our notation. We separate 3D momentum $\mathbf{k}$ in the BZ into $\mathbf{k}_{\perp}$ and $k_{\parallel}$ that are respectively perpendicular or parallel to the direction of the \emph{symmetry axis} (i.e., the rotation axis in 3D and/or mirror axis in 2D).  We call the subspace spanned by $\mathbf{k}_{\perp}$ at each given $k_{\parallel}$  the reduced Brillouin zone (rBZ). Then, we consider tight-binding Hamiltonians written in a basis $\{\ket{\phi_{l_{i}}(\mathbf{R})}\}$ where $\mathbf{R}$ represents the unit cell coordinates, and  $i=1,2,\ldots$ labels the orbitals with \emph{symmetry eigenvalues}  $l_{i}$ in each unit cell (i.e., eigenvalues of the rotation/mirror operator). We assume that all $\ket{\phi_{l_{i}}}$'s with different $l_{i}$ in one unit cell have their center localized on the same symmetry axis, with respect to which the symmetry eigenvalues are defined.

If a point in the rBZ remains invariant under a rotation/mirror transformation up to a reciprocal lattice vector, we then call this point a high symmetry point and denote it as $\boldsymbol{\Lambda}$. At each high symmetry point, we define the Berry phase of the Bloch bands in sector $l$ along the direction of the symmetry axis as
\begin{equation}
\label{eq:berryphase}
\gamma_{l}(\boldsymbol{\Lambda})=\int_{0}^{2\pi} dk_{\parallel}\sum_{n}i\inp{u^{l}_{n}(k_{\parallel},\boldsymbol{\Lambda})}{\partial_{k_{\parallel}}u^{l}_{n}(k_{\parallel},\boldsymbol{\Lambda})},
\end{equation}
where $\ket{u_{n}^{l}(k_{\parallel},\boldsymbol{\Lambda})}$ is the $n$-th eigenband with rotation/mirror eigenvalue $l$ at $\mathbf{k}=(k_{\parallel},\boldsymbol{\Lambda})$, and $i\inp{u^{l}_{n}(k_{\parallel},\boldsymbol{\Lambda})}{\partial_{k_{\parallel}}u^{l}_{n}(k_{\parallel},\boldsymbol{\Lambda})}\equiv\mathcal{A}_{l}(\mathbf{k})$ is the Berry connection, and $n$ is summed over the whole Hilbert space, i.e., occupied and unoccupied states. The Berry phase $\gamma_{l}(\boldsymbol{\Lambda})$ has the physical meaning of polarization along the symmetry axis contributed by all states with momentum $\boldsymbol{\Lambda}$ and symmetry eigenvalue $l$. We note that in this work we
always assume the Bloch Hamiltonian is periodic in momentum space, i.e., $H(\mathbf{k}+\mathbf{G})=H(\mathbf{k})$ where $\mathbf{G}$ is a reciprocal lattice vector. For simplicity we also ignore the internal structure of the unit cell, i.e., we assume all orbitals in one unit cell have the same position coordinates within the unit cell, which we take to be the position reference point. More detailed discussions which relax this assumption can be found in Ref.~\onlinecite{nelson:2021delicate}.

With $\gamma_{l}(\boldsymbol{\Lambda})$ defined in Eq.~\eqref{eq:berryphase}, let us now discuss the $2\pi$-quantized difference between $\gamma_{l}$ at a pair of high symmetry points in the rBZ. We denote the set  of symmetry eigenvalues of all occupied (unoccupied) bands at $\boldsymbol{\Lambda}$ as $l_{v}(\boldsymbol{\Lambda})$ ($l_{c}(\boldsymbol{\Lambda})$). If the intersection of $l_{v}(\boldsymbol{\Lambda})$ and $l_{c}(\boldsymbol{\Lambda})$ is empty, i.e., the mutually disjoint condition $l_{v}(\mathbf{\Lambda})\cap l_{c}(\mathbf{\Lambda})=\varnothing$ is satisfied, then: (i) the Berry phase $\gamma_{v}(\boldsymbol{\Lambda})$ of all occupied bands at a high symmetry point $\boldsymbol{\Lambda}$ can be expressed as $\gamma_{v}(\boldsymbol{\Lambda})=\sum_{l\in l_{v}}\gamma_{l}(\boldsymbol{\Lambda})$, and (ii) $\gamma_{l}(\boldsymbol{\Lambda})\equiv 2\pi n$ with $n$ an integer for any $l$. Point (i) is obviously true, so let us now explain point (ii). At a high symmetry point $\boldsymbol{\Lambda}$ in the rBZ, the Bloch Hamiltonian is block-diagonal. Each of its  blocks, $H_{l}(k_{\parallel},\boldsymbol{\Lambda})$, can be viewed as a 1D tight-binding Hamiltonian in the symmetry sector labeled by $l$, i.e., 
the Hilbert space spanned by all $\ket{\phi_{l_{i}}}$'s with $l_{i}=l$. The Berry phase of \emph{all} Bloch bands (i.e., both occupied and unoccupied Bloch bands) of such a 1D tight-binding model is always an integer multiple of $2\pi$, i.e., $2\pi n(\mathbf{\Lambda})$ where $n(\mathbf{\Lambda})$ is the integer multiple at $\mathbf{\Lambda}$. This is because the Wannier orbitals constructed by inverse-Fourier transforming all Bloch bands are just the basis orbitals centered at the reference point up to an integer number of lattice constants.  

After establishing points (i) and (ii), we can conclude that if at all $\mathbf{\Lambda}$'s, the mutually disjoint condition is satisfied, then the difference between two Berry phases at a pair of high symmetry points, i.e., $\gamma_{v}(\boldsymbol{\Lambda}_{1})-\gamma_{v}(\boldsymbol{\Lambda}_{2})$, should be quantized to $2\pi m$ with $m$ an integer. Given that the polarization along the symmetry axis should change continuously from $\boldsymbol{\Lambda}_{1}$ to $\boldsymbol{\Lambda}_{2}$ in the rBZ, the $2\pi m$ quantization of $\gamma_{v}(\boldsymbol{\Lambda}_{1})-\gamma_{v}(\boldsymbol{\Lambda}_{2})$ necessarily indicates an RTP if the system has vanishing Chern numbers, as will be the case for delicate topology\footnote{Remember that to discuss the delicate topology in a system, the nontrivial Chern number is precluded by assumption.}. Explicitly, the Berry phase difference indicates that $m$ charges are pumped toward one direction along the symmetry axis as we move from $\boldsymbol{\Lambda}_{1}$ to $\boldsymbol{\Lambda}_{2}$, and they are then pumped back when we complete the other half the loop in the rBZ from $\boldsymbol{\Lambda}_{2}$ to $\boldsymbol{\Lambda}_{1}$. In the following, we say the system has an RTP $m$ along the path $\boldsymbol{\Lambda}_{1}\to\boldsymbol{\Lambda}_{2}\to\boldsymbol{\Lambda}_{1}$ if $\gamma_{v}(\boldsymbol{\Lambda}_{1})-\gamma_{v}(\boldsymbol{\Lambda}_{2})=2\pi m$. The RTP is protected as long as the bulk energy gap, the rotation/mirror symmetry, and the Hilbert space restriction $l_{v}(\mathbf{\Lambda})\cap l_{c}(\mathbf{\Lambda})=\varnothing$ are preserved. This kind of topology is delicate in the sense that adding extra trivial bands (in either the occupied or unoccupied subspaces) that make the intersection of $l_{v}$ and $l_{c}$ at any $\mathbf{\Lambda}$ non-empty, can destabilize the RTP.

If we apply Stokes theorem we find
\begin{equation}
    \gamma_{v}(\boldsymbol{\Lambda}_{1})-\gamma_{v}(\boldsymbol{\Lambda}_{2})=\int d\mathbf{k}_{\parallel}\int_{\boldsymbol{\Lambda}_1}^{\boldsymbol{\Lambda}_2} dk_{\perp}\operatorname{Tr}\Omega_v(\mathbf{k}),
\end{equation}\noindent where $\Omega_{v}(\mathbf{k})$ is the Berry curvature in the occupied subspace, and the integral $dk_{\perp}$ is on a 1D path from $\boldsymbol{\Lambda}_{1}$ to $\boldsymbol{\Lambda}_{2}.$
Similarly, we can also define the Berry phase along the direction of the symmetry axis of all unoccupied bands as $\gamma_{c}(\boldsymbol{\Lambda})=\sum_{l\in l_{c}}\gamma_{l}(\boldsymbol{\Lambda})$. We denote the Berry curvature in the unoccupied subspace as $\Omega_{c}(\mathbf{k}).$ For any Bloch Hamiltonian $\operatorname{Tr}\Omega_{v}(\mathbf{k})=-\operatorname{Tr}\Omega_{c}(\mathbf{k})$, so we can conclude that the RTP in the unoccupied subspace is opposite to that in occupied subspace, i.e., $\gamma_{v}(\boldsymbol{\Lambda}_{1})-\gamma_{v}(\boldsymbol{\Lambda}_{2})=-(\gamma_{c}(\boldsymbol{\Lambda}_{1})-\gamma_{c}(\boldsymbol{\Lambda}_{2})).$ 

We also recall that delicate TIs with nonzero RTP generate gapless boundary modes on a sharp boundary. A sharp boundary means that the Hamiltonian of a system with boundaries is almost the same as the corresponding periodic Hamiltonian except that all hopping matrix elements across the boundaries are turned off. Intuitively, the bulk RTP implies that states with specific symmetry eigenvalues protrude from the bulk at the boundary, and these states must be ``compensated" by surface states with complementary symmetry eigenvalues to keep the balance of states in different symmetry sectors in each layer.  These boundary states are gapless on only a sharp boundary because the sharpness of the boundary constrains the energy of the surface states and forces them to cross the insulating bulk gap. This intuitive picture will become more transparent in the following example.

\begin{figure}[h]
\centering
\includegraphics[width=1\columnwidth]{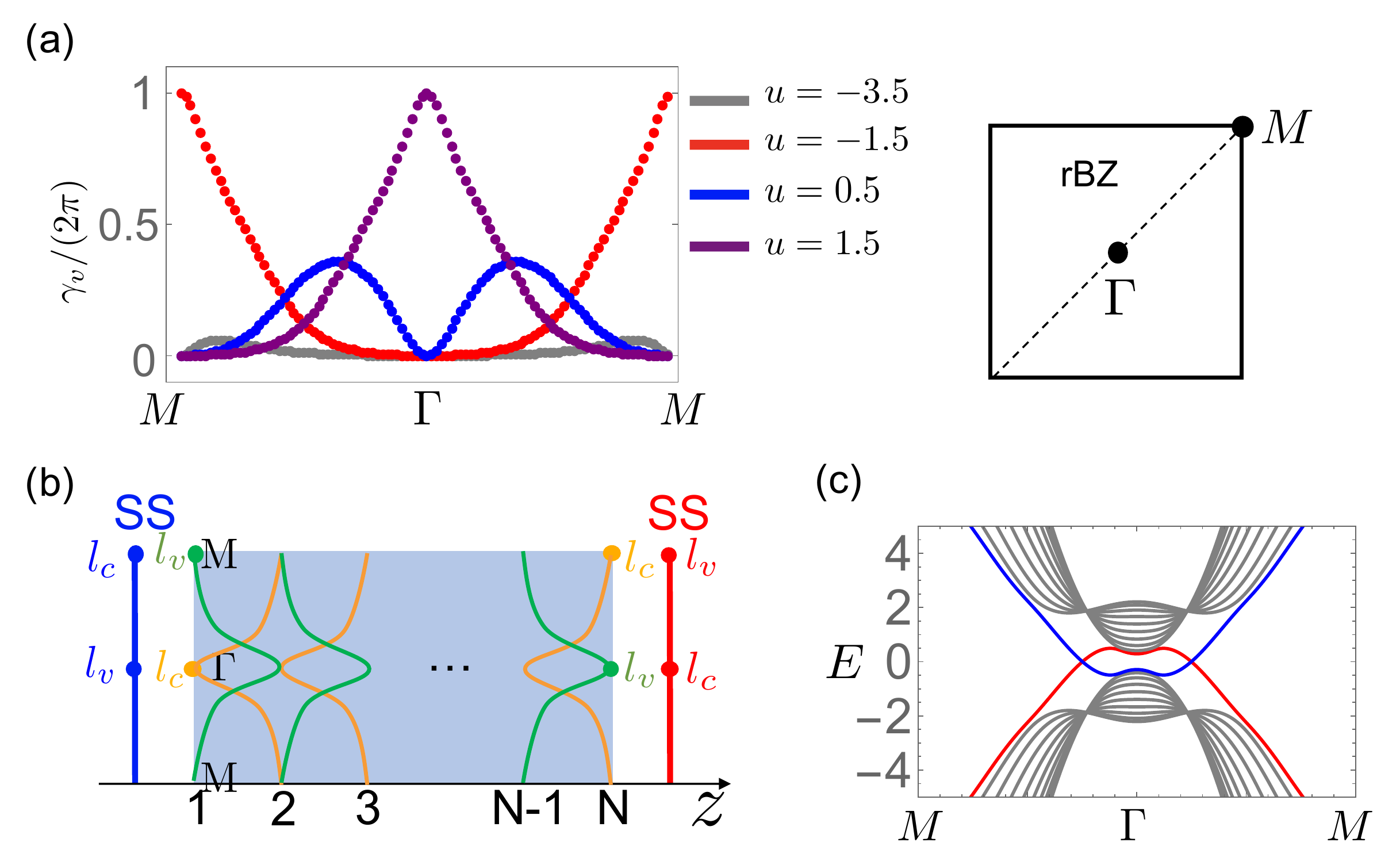}
\caption{The left panel of (a) shows the RTP (change of $\gamma_{v}/(2\pi)$) along the dotted-line path shown in the right panel. We include plots for the MRW model with $u=-3.5,-1.5,0,1.5$. The right panel of (a) shows $C_{4}$ rotation invariant points $\Gamma$ and $M$ in the rBZ.  (b) Illustration showing bulk states protruding into the surface (green (orange) color for occupied (unoccupied) states). The red (blue) line represents surface states localized on the top (bottom) surface with the compensating angular momentum at $\Gamma$ and $M$. (c) shows an exact diagonalization calculation of the gapless surface states of the MRW model at $u=1.5$ with 10 unit cells along the open $z$-direction (with sharp boundary), and periodic boundaries in $x,y$. The surface band localized on the top ($z=10$) and the bottom ($z=1$) surface are highlighted by red and blue colors, respectively. }
\label{fig:RTPhopf}
\end{figure}

To illustrate all the concepts and conclusions discussed above, let us take the $C_{4}$-symmetric two-band tight-binding model of the Hopf insulator constructed by Moore, Ran, and Wen (MRW) \cite{moore:2008topological}. The Hamiltonian of the MRW model is constructed from the well-known Hopf map:
\begin{equation}
\label{eq: MRW}
\begin{aligned}
&z=(z_1+iz_2,z_3+iz_4)^{T},
\\
&\mathbf{d}=z^{\dag}\bm{\sigma}z, \ \bm{\sigma}=(\sigma_{x},\sigma_{y},\sigma_{z}),
\\
&H_{\text{MRW}}(\mathbf{k})=\mathbf{d}\cdot\bm{\sigma},
\end{aligned}
\end{equation}
where $\sigma_{x},\sigma_{y},\sigma_{z}$ are Pauli matrices, and
\begin{equation}
\label{eq:z1234}
\begin{aligned}
&z_{1}=\sin k_{x}, z_{2}=\sin k_{y}, z_{3}=\sin k_{z},
\\
&z_{4}=u-\cos k_{x}-\cos k_{y}-\cos k_{z}.
\end{aligned}
\end{equation}
The Hamiltonian $H_{\text{MRW}}(\mathbf{k})$ has a $C_{4}$ rotation symmetry along the $z$-axis with rotation operator $C_{4z}=\exp\left(i\pi\sigma_{z}/4\right)$. For $u\neq\pm 1, \pm 3$, $H_{MRW}(\mathbf{k})$ is gapped and always has $l_{v}=\exp(i\pi/4)$ and $l_{c}=\exp(-i\pi/4)$ at both high symmetry points $\Gamma$ ($k_{x}=k_{y}=0$) and $M$ ($k_{x}=k_{y}=\pi$). This satisfies the mutually disjoint Hilbert space restriction discussed above. We find that $\gamma_{v}(\Gamma)$ and $\gamma_{v}(M)$ always have a $2\pi$-quantized difference when $1<|u|<3$ where the bulk Hopf invariant is $-1$ \footnote{When $|u|<1$, the bulk Hopf number is $2$ and $\gamma_{v}(\Gamma)-\gamma_{v}(M)=0$. However, there is a $2\pi$-quantized difference between $\gamma_{v}(\Gamma)$ and $\gamma_{v}(X)$ that is protected by the two-fold rotation symmetry. \cite{nelson:2021delicate}}, which indicates a non-vanishing RTP along $M\to\Gamma\to M$ as shown in Fig.~\ref{fig:RTPhopf} (a). 

Let us now focus on the MRW model with $u=1.5$ where $\gamma_{v}(\Gamma)-\gamma_{v}(M)=2\pi,$ and $\gamma_{c}(\Gamma)-\gamma_{c}(M)=-2\pi$. This RTP implies that in the bulk, the occupied (unoccupied) states at $\Gamma$  are one layer higher (lower) in the $z$-direction than the occupied (unoccupied) states at $M$, as illustrated by red (orange) lines in Fig.~\ref{fig:RTPhopf} (b).  Thus, on the top surface near $z=N$ there will be one extra occupied (unoccupied) state with $l=l_{v}$ ($l=l_{c}$) protruding from the bulk to the surface at $\Gamma$ ($M$). 
If we count both occupied and unoccupied states there should always be states with $l=l_{v}$ and $l_{c}$ at both $\Gamma$ and $M$ in each layer along the $z$-direction. Thus, the extra state (protruding from the bulk) with $l=l_{v}$ ($l=l_{c}$) at $\Gamma$ ($M$) in the surface layer should be ``compensated" by a surface state with $l=l_{c}$ at $\Gamma$ ($l=l_{v}$ at $M$). 

While the previous argument is generally true, the energies of the compensating surface states are not generically required to cross the energy gap, i.e., the states localized on the surface could only have energies in the bulk band regions. If we make a further assumption of a sharp boundary then we can make more definitive statements about surface state energies. The sharpness of the boundary guarantees that each diagonal block $H_{l}(\boldsymbol{\Lambda})$ of the effective 1D tight-binding Hamiltonian at $M$ or $\Gamma$ is a block Toeplitz matrix, i.e., the hopping matrix elements between unit cells at $\mathbf{R}$ and $\mathbf{R}^{\prime}$ depend on only $\mathbf{R}-\mathbf{R}^{\prime}$ and satisfy $H_{l,\mathbf{R}\mathbf{R}^{\prime}}\equiv H_{l,\mathbf{R}-\mathbf{R}^{\prime}}=(H_{l,\mathbf{R}^{\prime}-\mathbf{R}})^{\dag}$. The spectrum theorem of block Toeplitz matrices tells us that in the thermodynamic limit, the spectrum of $H_{l}(\boldsymbol{\Lambda})$ is bounded by the spectrum of its corresponding Bloch Hamiltonian $H_{l}(k_{z},\boldsymbol{\Lambda})$ (i.e., the corresponding Hamiltonian under periodic boundary conditions) \cite{nelson:2021delicate}. Thus, the surface state at $\Gamma$ ($M$) in the $l=l_{c}$ ($l=l_{v}$) sector must lie in the bulk 
unoccupied (occupied) band of the energy spectrum. This then guarantees that the surface bands must cross the bulk gap as $k_{\perp}$ goes from $\Gamma$ to $M$ shown in Fig.~\ref{fig:RTPhopf} (c). The same arguments can also be made for the bottom surface. In summary, if we relax the sharpness condition, the state-compensation argument is still valid, but the surface states are no longer guaranteed to have energies in the insulating gap unless additional symmetries are imposed as we show below.

\section{Scattering Theory for Delicate Topology}
\label{sec:pump}
\subsection{Pumped charge and reflection phase}
\label{sec:pump1}
From the review in Sec.~\ref{sec:review}, we see that  the RTP over a loop in the rBZ describes an adiabatic pump in an effective 1D system. During this pump, charge is pumped in one direction during part of the cycle, and then returns to its starting point at the end. To understand the relation between the RTP and the phase of a reflection amplitude in a scattering theory, we will first discuss the general correspondence between the reflection phase and the charge pumped in an adiabatically driven 1D wire. For simplicity, in this section, we denote the general adiabatic parameter as $\tau$, and assume the adiabatic period is from $\tau=0$ to $\tau=T$.

\begin{figure}[h]
\centering
\includegraphics[width=1\columnwidth]{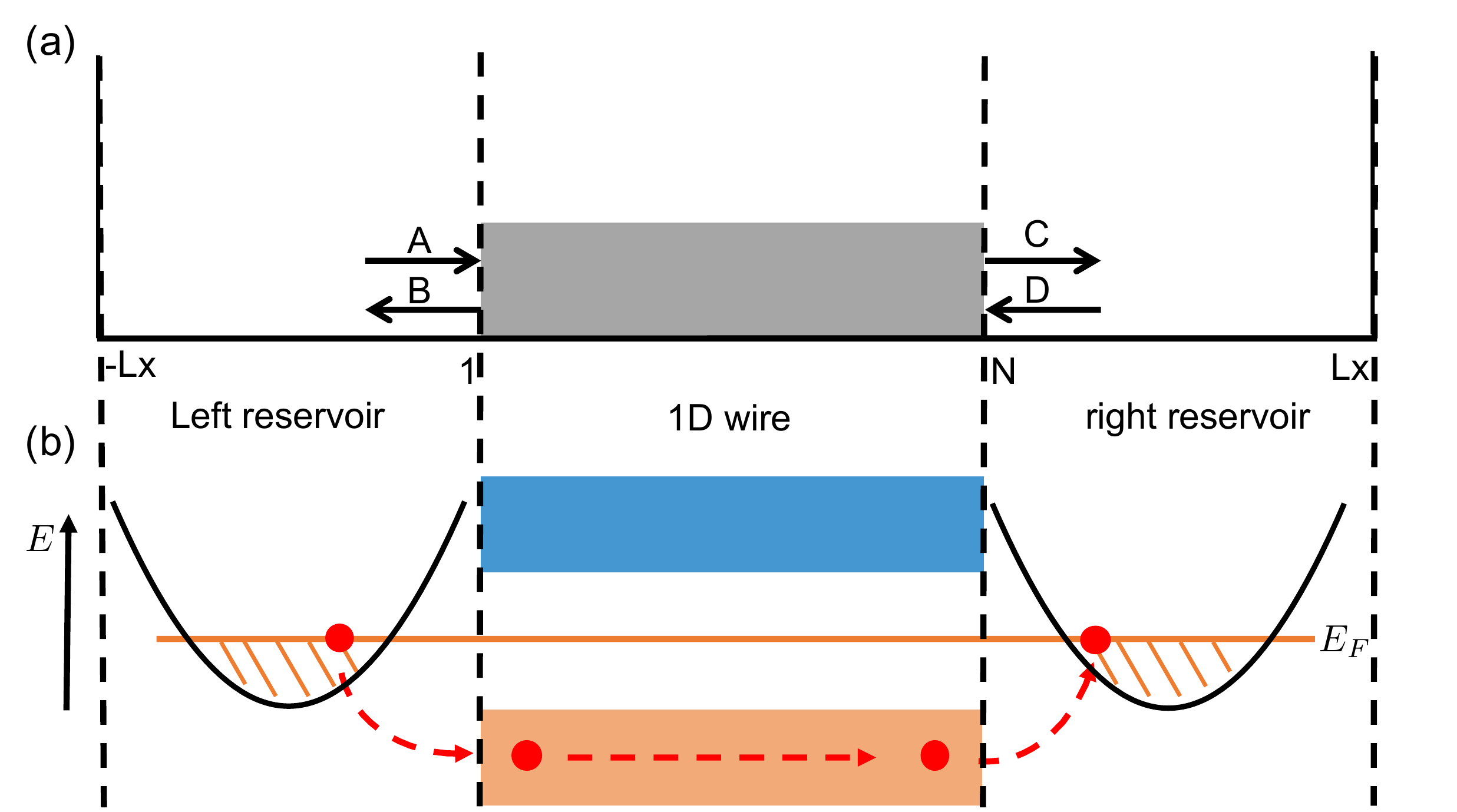}
\caption{(a)A schematic illustration showing the setup of the scattering process in a 1D geometry.  (b) An illustration showing the energy spectra of the left and right reservoirs and the central 1D wire. The orange horizontal line denotes the Fermi level $E_{F}$. The red dots represent charges, and the dotted red arrow indicates a pump of charges from the left reservoir to the right reservoir.}
\label{fig:scattering}
\end{figure}

As shown in Fig.~\ref{fig:scattering} (a), to study the scattering problem in a 1D geometry, we connect two reservoirs to a 1D wire; one on the left (L) and another on the right (R).  Fig.~\ref{fig:scattering} (a) shows incoming and outgoing waves where the wavefunctions in the reservoirs take the form
$\psi^{L}=Ae^{ik(x-x_{L})}+Be^{-ik(x-x_{L})}$ and $\psi^{R}=Ce^{ik(x-x_{R})}+De^{-ik(x-x_{R})}$, where superscripts $L$ and $R$ label the left and right reservoirs, and $x_{L}$ and $x_{R}$ are the reference points where the reflection and the transmission occur with the left and right reservoirs respectively. The incoming channels $A$ and $D$ are related to the outgoing channels $B$ and $C$ through the scattering matrix $S$:
\begin{equation}
\label{eq:S}
    S\begin{pmatrix}
    A
    \\
    D
    \end{pmatrix}\equiv \begin{pmatrix}
    r & t^{\prime}
    \\
    t & r^{\prime}
    \end{pmatrix}\begin{pmatrix}
    A
    \\
    D
    \end{pmatrix}=\begin{pmatrix}
    B
    \\
    C
    \end{pmatrix},
\end{equation}
where $r$ ($r^{\prime}$) and $t$ ($t^{\prime}$) are the reflection and transmission matrices of the left (right) interfaces. If the input waves have an energy that falls in a bulk energy gap of the wire in the central region, then there will be no bulk extended eigenstates that facilitate the transmission across the gapped wire. In this case, any transmission would come from only the overlap between possible boundary states that are exponentially localized on the two boundaries. However,  this process is exponentially suppressed as a function of the size of the central wire region\cite{braunlich:2010equivalence,fulga:2012scattering}. In such a scenario we therefore expect that in the thermodynamic limit the reflection matrices $r$ and $r^{\prime}$,  which respectively describe the reflection of input waves from the left and right interfaces, exponentially approach unit modulus, while the transmission matrices $t$ and $t^{\prime}$,  which respectively describe the transmission of input waves from the left and right interfaces, are exponentially suppressed. Generally, the wavefunctions in  both reservoirs can be written as $\psi^{L}=A(e^{ik(x-x_{L})}+re^{-ik(x-x_{L})})+t^{\prime}De^{-ik(x-x_{L})}$ and $\psi^{R}=D(r^{\prime}e^{ik(x-x_{R})}+e^{-ik(x-x_{R})})+t A e^{ik(x-x_{R})}$ with $k\geqslant 0$.  As argued above, $t$ and $t^{\prime}$ go to zero in the thermodynamic limit if the input waves have energy that falls in a bulk energy gap of the wire. In this scenario the wavefunctions in  both reservoirs become $\psi^{L}=A(e^{ik(x-x_{L})}+re^{-ik(x-x_{L})})$ and $\psi^{R}=D(r^{\prime}e^{ik(x-x_{R})}+e^{-ik(x-x_{R})}),$ where $r$ and $r^{\prime}$ are both unit modulus.

Let us now consider the reservoirs to be closed  (finite but large, i.e., $|L_{x}|\rightarrow +\infty$) boxes, i.e., any wavefunction in the left or right reservoirs satisfies $\psi^{R}(x=L_x)=\psi^{L}(x=-L_x)=0$. If we express the reflection for input waves with an energy in the bulk gap of the 1D wire as $r=e^{-i\phi}$ ($r^{\prime}=e^{-i\phi^{\prime}}$), the boundary conditions at $\pm L_{x}$ dictate that $k$ takes discrete values for a given $\phi$: 
\begin{equation}
\label{eq: discretek}
\begin{aligned}
&k^{L}_{n}=\frac{2n+1}{2(L_x+x_{L})}\pi+\frac{\phi}{2(L_x+x_{L})}, 
\\
&k^{R}_{n}=\frac{2n+1}{2(L_x-x_{R})}\pi+\frac{\phi^{\prime}}{2(L_x-x_{R})}.
\end{aligned}
\end{equation}

With this set up we will first review the scattering problem when the 1D wire represents a Thouless pump \cite{thouless1983} with Chern number one. During such a pumping process we imagine tuning the wire adiabatically over a 
period (i.e., from $\tau=0$ to $\tau=T$), such that one charge is pumped from the left reservoir to the right reservoir as indicated by the red dotted arrows in Fig.~\ref{fig:scattering} (b).  Since the wire returns to its initial state at $\tau=T,$ the reflection phases $\phi$ and $\phi^{\prime}$ can change only by integer multiples of $2\pi$ if they change continuously during this process. Thus $r$ should return to the same value it had at $\tau=0$. Indeed, previous work on the scattering formalism of the Thouless pump with unit Chern number has shown that if $\phi$ ($\phi^{\prime}$) changes continuously, then it must have a non-vanishing change of $2\pi$ ($-2\pi$) during the period \cite{fulga:2012scattering,braunlich:2010equivalence,PhysRevB.77.033304,franca2021non}. 

Let us try to understand the microscopic mechanism of this process. From Eq.~\eqref{eq: discretek} we can identify that the change of $\phi$ $(\phi^{\prime})$ generates a spectral flow such that $k^{L}_{n}\rightarrow k^{L}_{n+1}$ ($k^{R}_{n+1}\rightarrow k^{R}_{n}$). Hence, the left (right) reservoir loses (gains) one charge near the Fermi level, which must come through the assistance of the gapless boundary states of the central wire that appear as we tune $\tau$. The boundary states accommodate the charge transfer indicated in Fig.~\ref{fig:scattering} (b). Therefore, a $2\pi$ winding of $\phi$ is linked to the charge transfer and can be observed assuming $\phi$ changes continuously when we adiabatically pump the wire.

With the RTP in mind, let us
consider a scenario where we know one charge has been pumped from the left reservoir to the right reservoir (with the assistance of gapless boundary states) over an incomplete part of the adiabatic period. Since the period is incomplete, the wire has not yet returned to its initial state, and one cannot assert that $\phi$ and  $\phi^{\prime}$ must have changed by a multiple of $2\pi$. However, the left (right) reservoir still loses (gains) one charge, and hence $\phi$ ($\phi^{\prime}$) must exhibit some change during the incomplete cycle such that one charge enters the wire from the left reservoir, and one charge leaves the wire to enter the right reservoir. We will illustrate in the following that if one charge is continuously pumped from the left reservoir to the right reservoir, then $\phi$ ($\phi^{\prime}$) crosses $(2n+1)\pi$ from below (above) once for some value of $n$ (i.e., $r$ ($r^{\prime}$) must cross $-1$ counterclockwise (clockwise) once on the unit circle). Indeed, this behavior will be a signature of the RTP.

To see this explicitly let us first consider the left reservoir. We recall that the only valid $k$ are positive because we have identified $e^{ikx}$ as input waves from the left. Hence, the allowed $k$ at $\tau=0$ are labeled by the set of $n$ such that $2k^{L}_{n}(L_{x}+x_{L})=(2n+1)\pi+\phi\geq 0.$ From this we see that
if $\phi$ increases from a value less than $-(2n_0+1)\pi$ (for some $n_0$) to a value greater than $-(2n_0+1)\pi$ as we tune $\tau$, then $2 k_{n_0}^{L}(L_x+x_{L})=(2n_0+1)\pi+\phi$ becomes positive, and thus one more allowed state appears at the bottom of the energy spectrum. In the thermodynamic limit, the number of states under the Fermi level in the reservoir is uniquely determined by the Fermi level $\mu$, i.e., $N(E\le\mu)=\int_{0}^{\mu}\rho(E)dE$, where $\rho(E)$ is the density of states for free particles, and the Fermi level $\mu$ is fixed to lie in the bulk gap of the wire over the whole adiabatic process. Hence, one extra state at the bottom of the energy spectrum implies that one occupied state at the Fermi level is pushed beyond the Fermi level, and thus enters the wire from the left reservoir. 

Now let us consider the right reservoir. If a state from the wire enters the right reservoir, we need $\phi^{\prime}$ to decrease from a value greater than $-(2n_0^{\prime}+1)\pi$ to a value less than $-(2n_0^{\prime}+1)\pi$ (for some $n_0^{\prime}$), which will remove a state with wave vector $k_{n_{0}^{\prime}}^{R}$ at the bottom of the spectrum and thus make room at the Fermi level for the electron coming from the wire. 
In conclusion, we have shown that if one charge is continuously pumped from the left reservoir to the right reservoir, then $\phi$ ($\phi^{\prime}$) crosses $(2n+1)\pi$ ($(2n'+1)\pi$) from below (above) once for some value of $n$ ($n'$), i.e., $r$ ($r^{\prime}$) must cross $-1$ counterclockwise (clockwise) once on the unit circle.  

This conclusion can be generalized to multi-channel cases with $n$ pumped charges  from the left reservoir to the right reservoir, where $r$ is generally a matrix. For an input wave with energy in the bulk gap of the 1D wire, eigenvalues of the reflection matrix $r$ must cross $-1$ counterclockwise (clockwise) $n$ times when $n$ charges have been pumped from the left (right) reservoir to the right (left) reservoir \footnote{The mathematical proof of the correspondence between a Thouless pump and the reflection phase winding in Ref. \onlinecite{braunlich:2010equivalence} implies the same conclusion.}. Similar conclusions can also be drawn for the eigenvalues of $r^{\prime}$. Though our argument relies on the occupation and spectrum of the states in the reservoirs, the above conclusions drawn for the reflection phase can be tested by looking at the reflection of an input wave at a given energy in the bulk gap of only the wire without considering states at any other energy [see the calculation of reflection phase in Appendix~\ref{sec:reflectioncalculation}].

\subsection{Probing the delicate TI by reflection phase on a sharp boundary}
\label{sec: relectionRTP}
\begin{figure}[h]
\centering
\includegraphics[width=1\columnwidth]{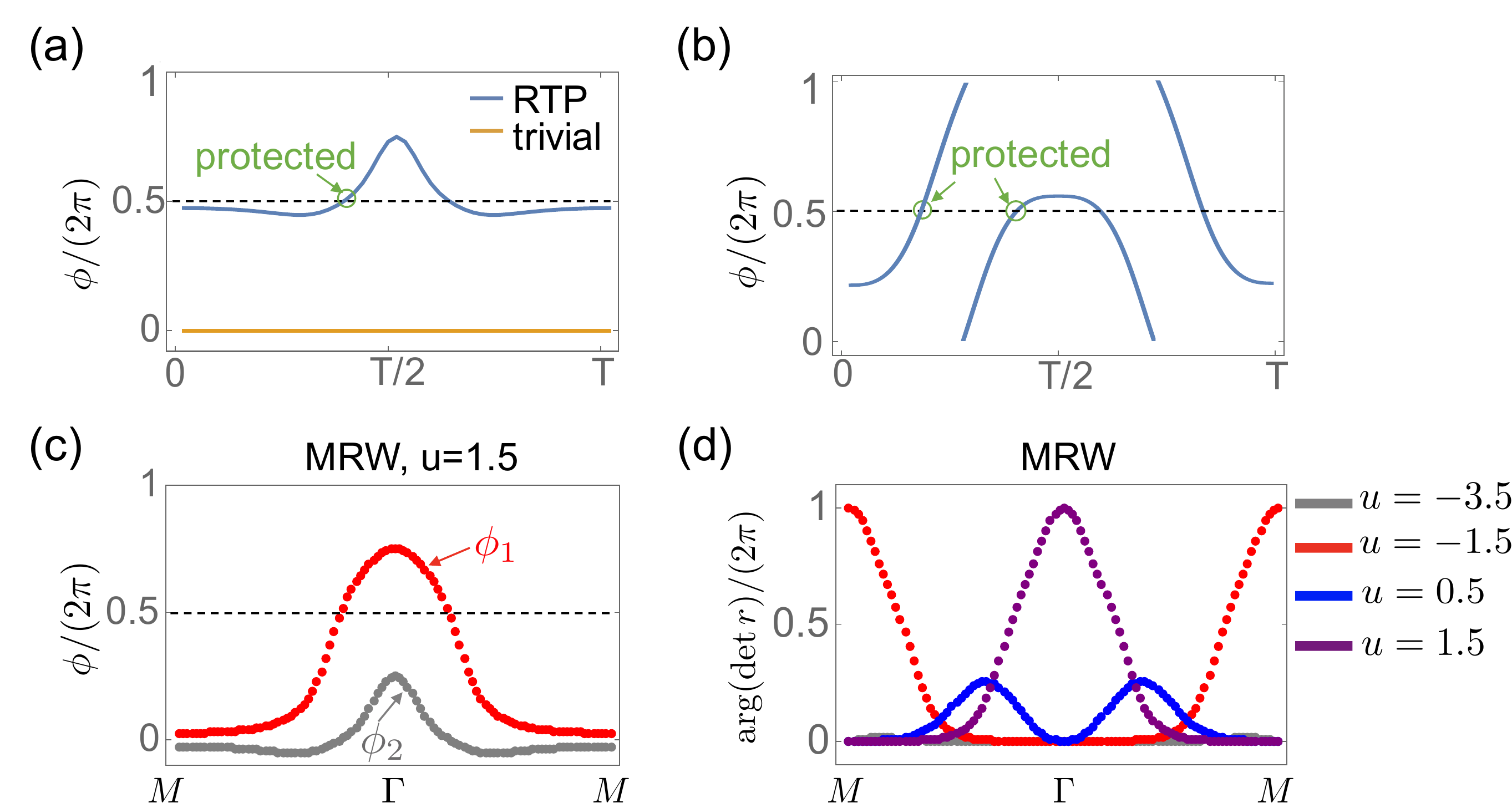}
\caption{(a) shows the crossing of the reflection phase $\phi$ with $\pi$. As discussed in the main text, the reflection phase $\phi$ of a delicate TI with RTP cannot be continuously deformed into the trivial case where $\phi=0$ for all $\tau$ while maintaining the bulk gap, symmetry, and sharp boundary. (b) illustrates the noncontractible structure of $\phi$ for a delicate TI with an RTP of two. 
The $x$-axis in (a) and (b) represent the adiabatic parameter, and $T$ is the period of the adiabatic cycle. (c) shows phases of two eigenvalues of the reflection matrix for the MRW model with $u=1.5$. (d) shows the returning winding of $\operatorname{arg}(\operatorname{det}r)$ for the MRW model with $u=-3.5, -1.5, 0.5, 1.5$.}
\label{fig:rphase}
\end{figure}

We can apply the above conclusions to the case of an RTP. We anticipate that an RTP can be detected by counting the net number of crossings of the eigenvalues of $r$ with $e^{i\pi}=-1$ during the first half of the adiabatic cycle from $\tau=0$ to $\tau=T/2$. However, this conclusion can be reached only if we input waves on a sharp boundary where there are gapless boundary modes (see Fig.~\ref{fig:RTPhopf} (c)) that assist the charge exchange between the reservoirs and the quantum wire. We need to also be sure to count the crossings with direction. Namely, if an eigenvalue crosses from below (counterclockwise in the complex plane) we count it as a positive crossing, otherwise it is a negative crossing. As discussed in Sec \ref{sec:pump1}, the net crossing over a half period for an insulator with sharp boundary is fixed by the number of charges pumped from the left reservoir to the right reservoir, i.e., the value of the RTP.  
 Hence, the net number of crossings of $\phi$ with $\pi$ cannot be removed by a continuous deformation of the Hamiltonian that preserves the bulk energy gap, the symmetry, and the sharp boundary. The existence of such protected crossings of the reflection phase distinguishes a non-trivial RTP from a trivial band insulator. For example, a trivial insulator that has a momentum-independent Hamiltonian will exhibit a $\phi$ fixed to $0$ during the whole period, 
as illustrated in Fig.~\ref{fig:rphase} (a).   This distinction is made more apparent when we have an RTP where $n>1.$ In this case the reflection phase has $n>1$ net crossings with $\pi$ over a half period. In Fig.~\ref{fig:rphase} (b) we plot $\phi$ versus $\tau$ for an example with $n=2.$ The figure clearly shows a non-contractible structure as $\tau$ goes from $0$ to $T/2$, i.e., $\phi$ first winds from $0$ to $2\pi$, and then keeps increasing enough to intersect $\pi$ again, but does not finish the second winding.

More explicitly, let us demonstrate that one can observe a returning winding in the phase of $\operatorname{det} r$ for two-band delicate TIs protected by rotation/mirror symmetry. In order to have delicate topology we want the two basis orbitals in each unit cell to have different symmetry eigenvalues. In a basis where the rotation/ mirror operator is diagonal we want the rotation/mirror generator to be $\sigma_{z}$. Then, at high symmetry points $\boldsymbol{\Lambda}$ in the rBZ, $H(k_{\parallel},\boldsymbol{\Lambda})$ is proportional to $\sigma_{z}$, which has an emergent chiral symmetry $\mathcal{C}=\sigma_{x}$. 

The emergent chiral symmetry will enforce the reflection matrix $r$ to be unitarily similar to $r^{\dag}$ \cite{fulga:2011scattering, fulga:2012scattering}. Thus, one can find a basis of incoming and outgoing states such that all eigenvalues of $r$ are real (i.e., $\pm1$). In such a basis,
previous work on the reflection matrices of chiral symmetric topological phases shows that the absolute value of the chiral winding number is equal to the number of $-1$ eigenvalues of $r$ \cite{fulga:2011scattering, fulga:2012scattering}. Since $H(k_{\parallel},\boldsymbol{\Lambda})\propto \sigma_{z}$ always has chiral winding number zero, we can conclude that all eigenvalues of $r$ at rotation/mirror invariant $\boldsymbol{\Lambda}$ in the rBZ must be $1$ in the basis where all eigenvalues of $r$ are real. Hence, the phase $\phi$ of each eigenvalue can only continuously change by integer multiples of $2\pi$ over an adiabatic path connecting a pair of rotation/mirror invariant $\boldsymbol{\Lambda}$. With this constraint, if the reflection phase changes continuously and crosses $\pm \pi$ $m$ times when required by a nonzero bulk RTP with value $m$, it is straightforward to see that the phase of $\operatorname{det}r$, $\operatorname{arg}(\operatorname{det}r)$ is basis-independent, and has a net winding  of $2m\pi$ over half a period. In the second half period, there will be a corresponding returning winding. Thus, for any 2-band delicate TIs, $\operatorname{arg}(\operatorname{det}r)$ is expected to show a returning winding that replicates the RTP. 

As an example, let us consider the 2-band MRW model with $u=1.5$ and $10$ unit cells along the $z$-direction. We identify $\tau=0,T$ with the $M$-point in the rBZ, and identify $\tau=T/2$ with the $\Gamma$-point in the rBZ. Thus, the adiabatic pump from $\tau=0$ to $\tau=T$ is now identified with the dotted line indicated in Fig.~\ref{fig:RTPhopf} (a). From $M$ (i.e., $\tau=0$) to $\Gamma$ (i.e., $\tau=T/2$), one charge is transferred from the left reservoir to the quantum wire with the assistance of gapless boundary states (see Fig.~\ref{fig:RTPhopf} (c)) localized on the boundary at $z=1.$ Additionally, one charge is transferred from the quantum wire to the right reservoir with the assistance of gapless boundary states (see Fig.~\ref{fig:RTPhopf} (c)) localized on the boundary at $z=10$. To confirm our predictions we numerically calculate the two-channel reflection matrix for a grid of values on the $M-\Gamma-M$ line, and plot the phases of its two eigenvalues in Fig.~\ref{fig:rphase} (c). As expected, we see one of the two phases, $\phi_{1}$, crosses $\pi$ once in the first half period, and once in the second half-period which captures the bulk RTP. We note that at $\Gamma$, $\phi_{1}$ and $\phi_{2}$ are not real because the basis we use for input and output waves in the numerical calculation is not the basis that makes $r=r^{\dag}$. However, $\phi_{1}=-\phi_{2} \ \operatorname{mod} \ 2\pi$, which is consistent with the fact that $r$ is unitarily similar to $r^{\dag}$ for a chiral symmetric 1D chain. Furthermore, we plot $\operatorname{arg}(\operatorname{det}r)$ over the $M-\Gamma-M$ line in the rBZ in Fig.~\ref{fig:rphase} (d). We see that the returning winding of $\operatorname{arg}(\operatorname{det}r)$ exactly duplicates the Berry phase calculations of the RTPs shown in Fig.~\ref{fig:RTPhopf} (a). The details about calculations of the reflection matrix for 1D chains are given in Appendix~\ref{sec:reflectioncalculation}.

\subsection{Beyond sharp boundaries}
If we relax the sharpness requirement of the boundaries, the boundary modes can be gapped out by adding local potentials on the boundary. In such a scenario  there will no longer be charge exchanges between the reservoirs and the quantum wire. For example, by adding local potentials on the boundary  that decrease (increase) the energy of the red (blue) color boundary band in Fig.~\ref{fig:RTPhopf} 
one can gap out the surface states as shown in Fig.~\ref{fig:gboundary} (a) \footnote{Details about the local potential can be found in Appendix~\ref{sec:boundarypotential}.}. Since the surface states no longer cross the Fermi level (of the quantum wire or the reservoirs), there cannot be charge exchange between the reservoir and the quantum wire, and thus the reflection phase will not cross $\pi$ anymore as we confirm in Fig.~\ref{fig:gboundary} (b). Instead, the pumped charge during the first half-period accumulates on the boundary at $z=10$ and polarizes the system along the $z$-direction, as shown in Fig.~\ref{fig:gboundary} (c)

\begin{figure}[h]
\centering
\includegraphics[width=1\columnwidth]{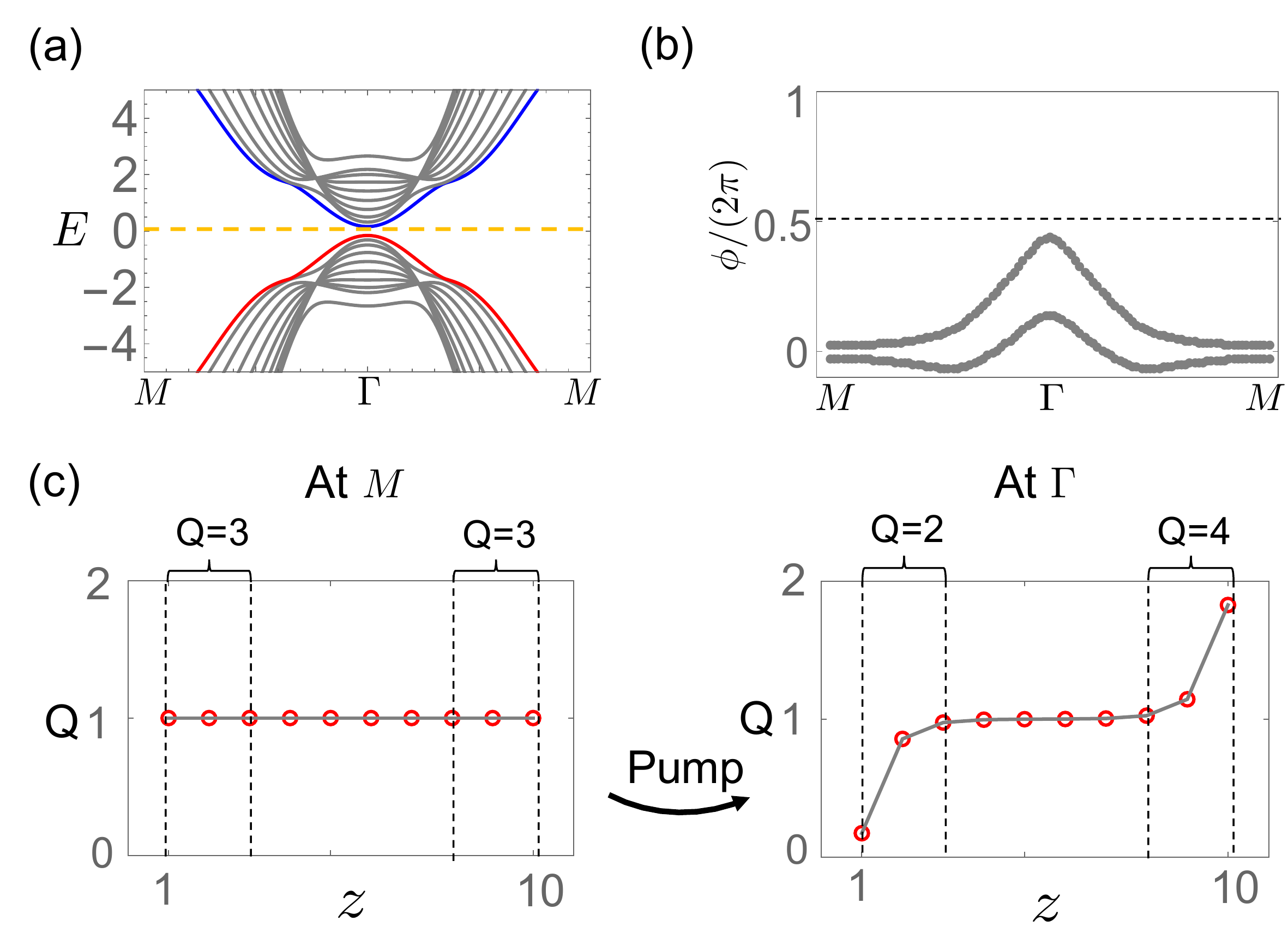}
\caption{(a) Gapped surface states of the MRW model when the boundary is decorated with a local potential. The MRW model has 10 unit cells along the $z$-direction and $u=1.5$. (b) shows the corresponding reflection phases that no longer cross $\pi$ since the boundary states are gapped. (c) shows the change of the charge distribution during the first half of the adiabatic cycle at $M$ (left) and $\Gamma$ (right).  At $M$, each unit cell has one charge and the combined charge in the 3 unit cells nearest each boundary is $Q=3$. At $\Gamma$, by looking at the total charge $Q$ on the three unit cells closest to the bottom boundary at $z=1$ (the top boundary at $z=10$), one can observe that there is exactly one less (more) charge on the bottom (top) boundary comparing to that at $M$. }
\label{fig:gboundary}
\end{figure}

The charge build-up mechanism that breaks the connection between the RTP and the reflection phase winding gives us a key hint about how to restore it. Indeed, to guarantee the nontrivial behavior of the reflection matrix discussed in Sec.~\ref{sec: relectionRTP}, we just need to make sure that the charge exchange between the reservoirs and the quantum wire is always allowed and that the system cannot polarize. 
 In general, extra symmetries can forbid the ground state of the quantum wire at $\tau=0$ and $\tau=T/2$ from having a polarized charge distribution like that in Fig.~\ref{fig:gboundary} (c). As a result, if we impose such symmetries the charge exchange between the reservoirs and the quantum wire would be guaranteed when there is a nonzero RTP in the quantum wire.

To implement the above requirement on the charge distribution, one can impose symmetries on the 1D Hamiltonians at the high symmetry points in the rBZ, i.e., impose symmetries at $\tau=0$ and $\tau=T/2$ in the adiabatic pump language. One choice is the requirement of chiral or particle-hole symmetry at high symmetry points in the BZ. In fact, for an insulator with a unity-valued RTP and chiral symmetry at $\tau=0$ and $\tau=T/2$ we know:
(i) the unoccupied subspace has an opposite RTP to the occupied subspace, i.e., if in the occupied subspace one charge accumulates on the right boundary, then in the unoccupied subspace, one charge should accumulate on the left, and (ii) in each unit cell, the chiral symmetry enforces the charge distribution of the occupied states to be the same as that of the unoccupied states. It is straightforward to see that (i) and (ii) can be simultaneously satisfied only if there is no charge accumulation on the boundary. Then, the pumped charges during the first half-period must go into the reservoir with the help of gapless boundary modes. One can also straightforwardly argue that particle-hole symmetry has the same effect. 

\begin{figure}[h]
\centering
\includegraphics[width=1\columnwidth]{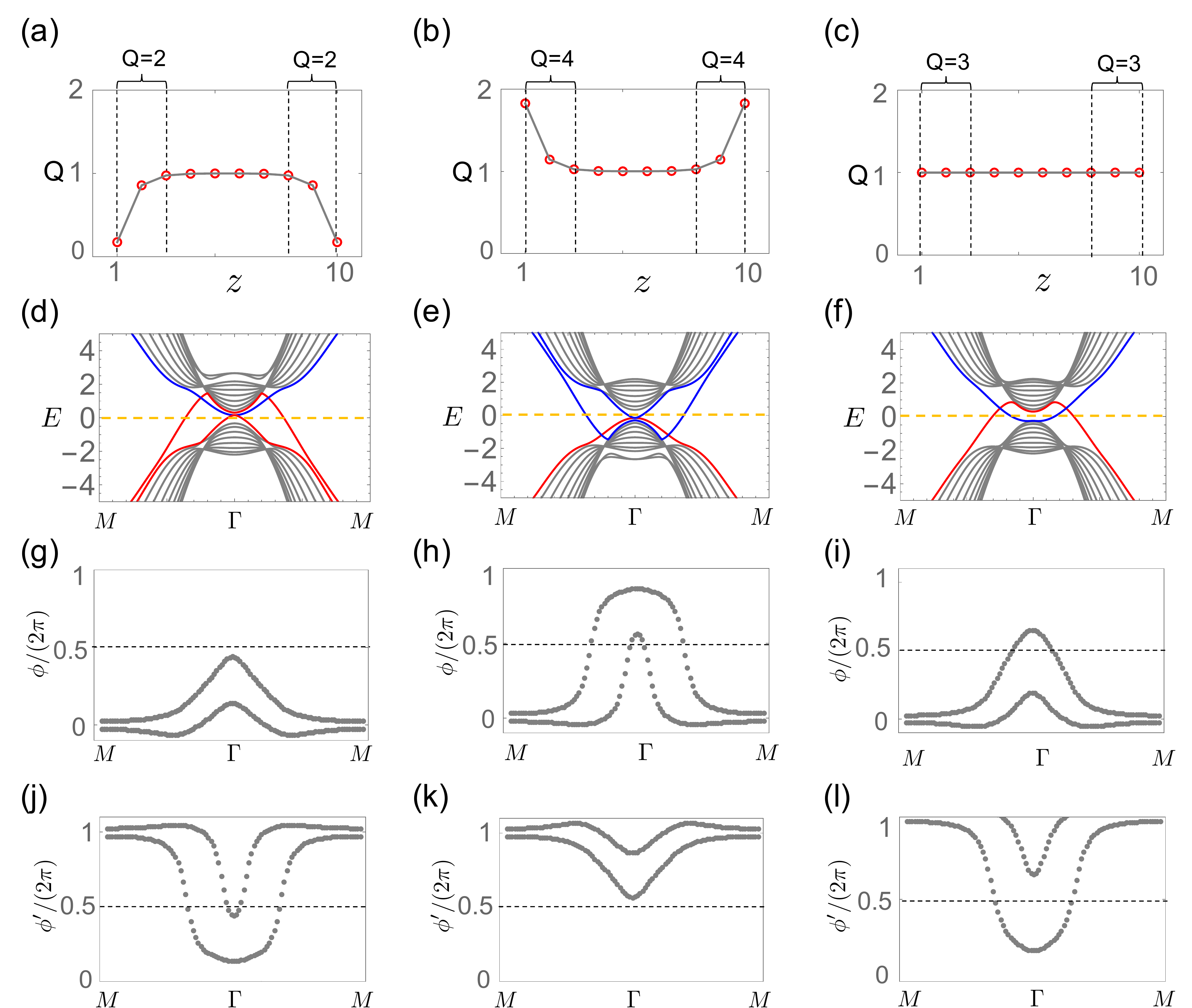}
\caption{The three columns of sub-figures represent calculations for three different boundary terminations that preserve mirror symmetry at the high symmetry points of the MRW model. The calculations are for $u=1.5$, and there are 10 unit cells along the $z$-direction. (a), (b), (c) show the charge distribution of the occupied ground state at ${\textbf{k}}=\Gamma$. Each unit cell has two sites/orbitals, and the summed charge density $Q$ in the last three unit cells near each of $z=1, 10$ are marked by the gray, dotted boxes. In (a)/(b), we see one more/less charge on \emph{both} boundaries compared to the cases where the boundary is sharp as in (c). The second row (d), (e), (f) shows gapless surface states localized near boundaries at $z=1$ (blue) and $z=10$ (red). The third row and the fourth row show the nontrivial crossings with $\pi$ of the reflection phases. The reflection phase $\phi$ at the $z=1$ boundary is shown in (g), (h), (i).  The reflection phase $\phi^{\prime}$ at the $z=10$ boundary are shown in (j),(k),(l).}
\label{fig:mboundary}
\end{figure}

Another, perhaps more realistic choice, is to require an additional mirror symmetry along the direction of the symmetry axis. With such a choice, the charge distribution in the quantum wire at high symmetry points (i.e., $\tau=0$ and $\tau=T/2$) must be mirror symmetric along the 1D chain,  and thus guarantees that the system will not be polarized and that there will be charge exchange between the reservoirs and a quantum wire with a nonzero RTP. As will be discussed below, the existence of charge exchange between the reservoirs and the quantum wire has an additional important consequence that gapless modes survive on the boundary of the quantum wire despite the boundary not being sharp.

Let us illustrate this idea for the MRW model at $u=1.5$  with 10 unit cells along the $z$-direction. Consider local perturbations on the boundary that are much smaller than the gap at $M,$ but comparable with the gap at $\Gamma$ \footnote{Details about the local potentials we use can be found in Appendix~\ref{sec:boundarypotential}.}. Then at $M$, we expect the charge distribution of the ground state to always be uniform as in the case without the perturbations. On the path from $M$ to $\Gamma$, the bulk, unity-valued RTP indicates that one charge is pumped from the boundary at $z=1$ to the boundary at $z=10$ in the wire. If the mirror symmetry is preserved during this process, then at $\Gamma$ there are three possible cases. 

(i) For the first scenario, two charges enter the reservoirs through the boundary at $z=10$. As shown in Fig.~\ref{fig:mboundary} (a), this leads to a mirror symmetric charge distribution at $\Gamma$  such that each boundary has one less charge compared to that at $M$ (which has a charge distribution identical to that in Fig. \ref{fig:mboundary}(c)). This also implies that there are two branches of surface bands near $z=10$ that cross the Fermi-level between $M$ and $\Gamma$ such that there are more states occupied at $M$ than $\Gamma$ (see Fig. \ref{fig:mboundary}(d)). The spectral flow from the two branches of surface states indicates that  there are two charges entering the right reservoir.   According to the theory discussed in Sec \ref{sec:pump1}, the two additional charges in the right reservoir after a half period indicate that the reflection phase near the $z=10$ boundary, $\phi^{\prime}$, would cross $\pi$ twice from above as shown in Fig. \ref{fig:mboundary}(j). However, the reflection phase near the $z=1$ boundary, $\phi$, does not cross $\pi$ as observed in Fig. \ref{fig:mboundary}(g). 

(ii) In a second scenario two charges enter the wire through the boundary at $z=1$, as shown in Fig.~\ref{fig:mboundary}(b). This leads to a mirror symmetric charge distribution at $\Gamma$ such that each boundary has one more charge compared with that at $M$.  Similarly, this also implies that from $M$ to $\Gamma$ there are two surface bands localized near $z=1$ that cross the Fermi level such that there are less occupied states at $M$ than are at $\Gamma$ (see Fig. \ref{fig:mboundary}(e)).  According to the theory discussed in Sec \ref{sec:pump1}, the reflection phase at the $z=1$ boundary, $\phi$, should cross $\pi$ twice from below (as shown in Fig. \ref{fig:mboundary}(h)) to indicate the left reservoir having two fewer charges. In comparison, the reflection phase near the $z=10$ boundary, $\phi^{\prime}$, does not cross $\pi$ as observed in Fig. \ref{fig:mboundary}(k).

(iii) For the third scenario, one charge enters the wire through the boundary at $z=1$, and one leaves the wire through the boundary at $z=10$, which is similar to the case when the boundary is sharp (see  Fig.~\ref{fig:mboundary}(c)). Here there is just one branch of edge states at $z=1$ and one branch at $z=10$ that cross the Fermi level between $M$ and $\Gamma$ (see Fig. \ref{fig:mboundary}(f)). Each of the reflection phases $\phi, \phi^{\prime}$ will exhibit a single crossing with $\pi$ (see Fig. \ref{fig:mboundary}(i)(l) respectively).   

As we can see, for all three cases, there are always gapless boundary states (albeit with different configurations) and we observe the predicted non-trivial behavior of the reflection phase crossings. 

After the above discussions, we close this section by emphasizing that for a system with boundaries that are not sharp, an imposed symmetry can help us have an observable RTP. Indeed, using the data in the reflection phase and the charge distribution, one can unambiguously determine the value of the RTP in the wire: the crossing of the reflection phase with $\pi$ over a half period provides the information about charge exchange between the wire and the reservoirs. Together with the charge distribution at $\tau=0$ and $\tau=T/2$, one can determine the number of charges pumped from one side of the wire to the other side of the wire over the half period. As an example for scenario (i) above, the reflection phase $\phi^{\prime}$ has two $\pi$-crossings which implies that two states are transferred to the right reservoir. If there is no pump then we would expect the charge near the right boundary to have two less charges than the initial case, but we see in Fig. \ref{fig:mboundary}(a) that it has only one less charge. No charges traveled into the left reservoir, but we see that the left edge also has one less charge. Thus, we can conclude that one charge was also pumped from left to right during this process, and hence the RTP has a value of unity. Finally, we emphasize that the existence of gapless boundary states in each scenario illustrates that imposing additional symmetry to a delicate topological insulator can stabilize boundary modes away from a sharp boundary limit.

\begin{figure*}[t]
\centering
\includegraphics[width=2\columnwidth]{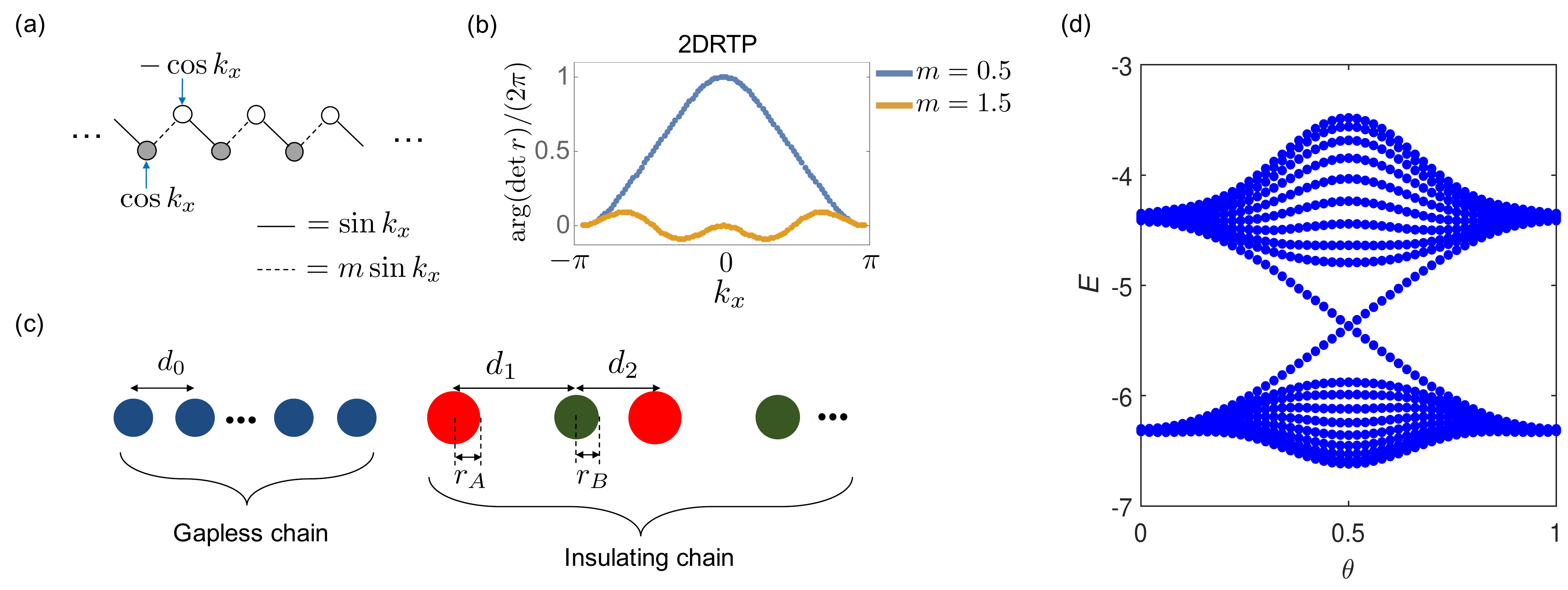}
\caption{(a) Illustration of a lattice model with the position space hoppings and staggered onsite potentials of $H_{\text{2DRT}}$ at a fixed $k_{x}$. (b) The returning winding of $\operatorname{arg}(\operatorname{det}r)$ for the 2D RTP model with $m=0.5,1.5$. (c) illustration of waveguides in a photonic crystal, where the disks with different colors represent waveguides with different radii. $d_{0,1,2}$ and $r_{A,B}$ are tuned by a parameter $\theta$ that can be identified with $k_{x}$ in $H_{\text{2DRT}}$. (d) Spectra of full-continuum waveguide chains with $\theta$ ranging from $0$ to $1$, which corresponds to $k_{x}$ in the range $0$ to $\pi$. Note that all energy scales are normalized with respect to the inter-cell 
hopping at $\theta=\pi/2$ in the insulating chain. }
\label{fig:2DRTP}
\end{figure*}

\section{Experimental proposal}
\label{sec:expproposal}

In this section, we provide an experimental proposal to implement a delicate TI in a photonic crystal. Furthermore, we demonstrate that the reflection phase can be measured from the wavefunction of static states in reservoirs. Since the hoppings in the MRW model are complicated to engineer in photonics, we instead study a 2D mirror symmetric model that also has a bulk RTP, but is more experimentally feasible. Let us consider a two-band, 2D Bloch Hamiltonian:
\begin{equation}
\label{eq:2DRTP}
\begin{aligned}
H_{\text{2DRTP}}(k_{x},k_{y})&=\sin(k_{x})(m+\cos( k_{y}))\sigma_{x}
\\
&+\sin(k_{x})\sin( k_{y})\sigma_{y}+\cos (k_{x})\sigma_{z},
\end{aligned}
\end{equation}
which has a mirror symmetry $m_x=\sigma_{z}$ along the $x$-direction,  and a time-reversal symmetry $\mathcal{T}=\sigma_{z}K$ where $K$ is the complex conjugation operator. When $0<m<1$, this model has a $-2\pi$ Berry flux on the $k_{x}>0$ half of the BZ, and thus has a $2\pi$ Berry flux on the $k_{x}<0$ half of the BZ to satisfy time-reversal symmetry. This configuration guarantees $\gamma_{v}(k_{x=0})-\gamma_{v}(k_{x=\pi})=2\pi$ by Stokes Theorem. Hence this model manifests a RTP \cite{alex:2022topological}. 

Treating $k_{x}$ as an adiabatic parameter, $H_{\text{2DRTP}}$ at each fixed $k_{x}$ is a Su-Schrieffer-Heeger(SSH)-type 1D chain along the $y$-direction with a staggered potential $\pm \cos k_{x}$ (see Fig. ~\ref{fig:2DRTP} (a)). We numerically calculate the single-channel reflection coefficient $r$ and extract  $\phi=\operatorname{arg}(r)$ for this 2D RTP model as shown in Fig.~\ref{fig:2DRTP} (b). The details of the single-channel calculations are discussed in Appendix ~\ref{sec:reflectioncalculation}.

Treating the 2D model in Eq.~\eqref{eq:2DRTP} as a set of 1D chains along the $y$-direction that are parameterized by different $k_{x}$ values, we can implement the model using 1D photonic crystals composed of evanescently coupled waveguides. The on-site potential and hopping strength depend on the radii of the waveguides (e.g., $r_{A}$ and $r_{B}$ in Fig.~\ref{fig:2DRTP} (c)), and the distance between the adjacent waveguides (e.g., $d_{0}$, $d_{1}$, and $d_{2}$ in Fig.~\ref{fig:2DRTP} (c)), respectively. These quantities can be fully controlled by fabricating the evanescently coupled waveguide array using 3D printing by two-photon lithography\cite{schulz2021topological,schulz2022photonic}.  The diffraction of light through the waveguide array is governed by the paraxial wave equation
\begin{align}
i\partial_{z}\psi(\textbf{r},z) = &\left[-\frac{1}{2k_{0}}\nabla^{2}_{\textbf{r}}-\frac{k_{0}\Delta n(\textbf{r},z)}{n_{0}}\right]\psi(\textbf{r},z) \nonumber \\ 
&\equiv H_{\textrm{cont}}\psi(\textbf{r},z),
\label{eq:paraxial}
\end{align}
where ${\textbf{r}}=(x,y)$, $\psi(\textbf{r},z)$ is the envelope function of the electric field $\textbf{E}(\textbf{r},z)=\psi(\textbf{r},z)e^{i(k_{0}z-\omega t)}\hat{x}$, $k_{0}=2\pi n_{0}/\lambda$ is the wave number within the medium, and $\Delta n$ is the refractive index of the waveguide relative to the index of our medium: $n_{0}$.
In addition, $\lambda$ is the wavelength of laser light, $\nabla^{2}_{\textbf{r}}$ is the Laplacian in the transverse $(x,y)$ plane, $\omega=2\pi c/\lambda$, and $H_{\textrm{cont}}$ is the continuum Hamiltonian for the propagation of light in the waveguide array.
The eigenvalues and eigenmodes of the waveguide array can be obtained by diagonalizing $H_{\textrm{cont}}$, which is called a full-continuum simulation \cite{NatPhoton7Rechtsman}.

Using realistic experimental values, we can parameterize a family of photonic waveguides by a variable $\theta$ via: $r_{A,B}(\theta)=1.5\pm 0.1\cos(\theta\pi)~\mu$m, $d_{0}=6.5~\mu$m, $d_{1}(\theta)=9-2.85\sin(\theta\pi)~\mu$m, $d_{2}(\theta)=8-3.0\sin(\theta\pi)~\mu$m.  
Through the full-continuum simulation, we confirmed that
we can tune $\theta,$ which controls the size of, and the distance between, waveguides, so that as it is tuned from $0$ to $1$ the system changes in the same manner that tuning $k_{x} \in [0,\pi]$ changes parameters of the tight-binding model in Eq. \ref{eq:2DRTP}. The energy spectrum of such a system derived from the continuum calculation is shown in Fig.~\ref{fig:2DRTP} (d), where we indeed observe the expected in-gap states in the first half period, i.e.,  from $\theta=0$ to $\theta=1,$ which corresponds to $k_{x}=0$ to $k_{x}=\pi$.  Here we show the spectrum of only the first half period, because with mirror symmetry $m_{x}$ we know that the spectrum of the second half period at a $k_{x0}\in[\pi,2\pi]$ is the same as the spectrum at $2\pi-k_{x0} \in [0,\pi]$.

To detect the reflection phase, we attach a 1D gapless chain $H_{\text{gapless}}=\sum_{y}h_{0} (c^{\dag}_{y+1}c_{y}+h.c.)$ to the left of each 1D insulating chain as shown in Fig.~\ref{fig:2DRTP} (c).  The attached gapless chain plays the role of a reservoir. For clarity, we label the central 1D insulating chain with the range  $ 1\leqslant y\leqslant L_0$, and label the gapless chain with the range $y\leqslant 0$. In our tight-binding calculations, we connect the gapless chain and the insulating chain between the sites $y=0$ and $y=1$ using a hopping with strength $\sin^2 k_{x}$. The effects of alternative choices of the form of this hopping are discussed in Appendix~\ref{sec:reflectioncalculation}. 

To set up the scattering problem let us consider an incoming wave $\psi_{\text{in}} (y)=e^{i k y}$ ($y=0,-1,-2,\ldots$) from the gapless chain directed toward the insulating chain with energy $E=0$ (i.e., the in-gap Fermi level of the insulating chain), where $y$ is the site coordinate, and $k>0$ is the wave vector. Note that for the gapless chain we use, an input wave with $E=0$ has a wave vector $k= \pi/2$ because the dispersion of the gapless chain is $E(k)=2h_{0}\cos k$. After a reflection by the insulating chain, there will be an outgoing wave with wave function $\psi_{\text{out}}(y)=e^{-i k y+\phi}$ as discussed above. Then, the wavefunction of the state with energy $E=0$ in the gapless reservoir chain is $\psi(y)=(e^{i k y}+e^{-i k y+\phi})/\sqrt{2}$. Thus, up to a normalization factor the intensity is given by
\begin{equation}
\label{eq:density}
|\psi(y)|^2=1+\cos(2 k y-\phi), 
\end{equation}
which contains information about $\phi$ and can be measured in the experiment.  

\begin{figure}[t]
\centering
\includegraphics[width=1\columnwidth]{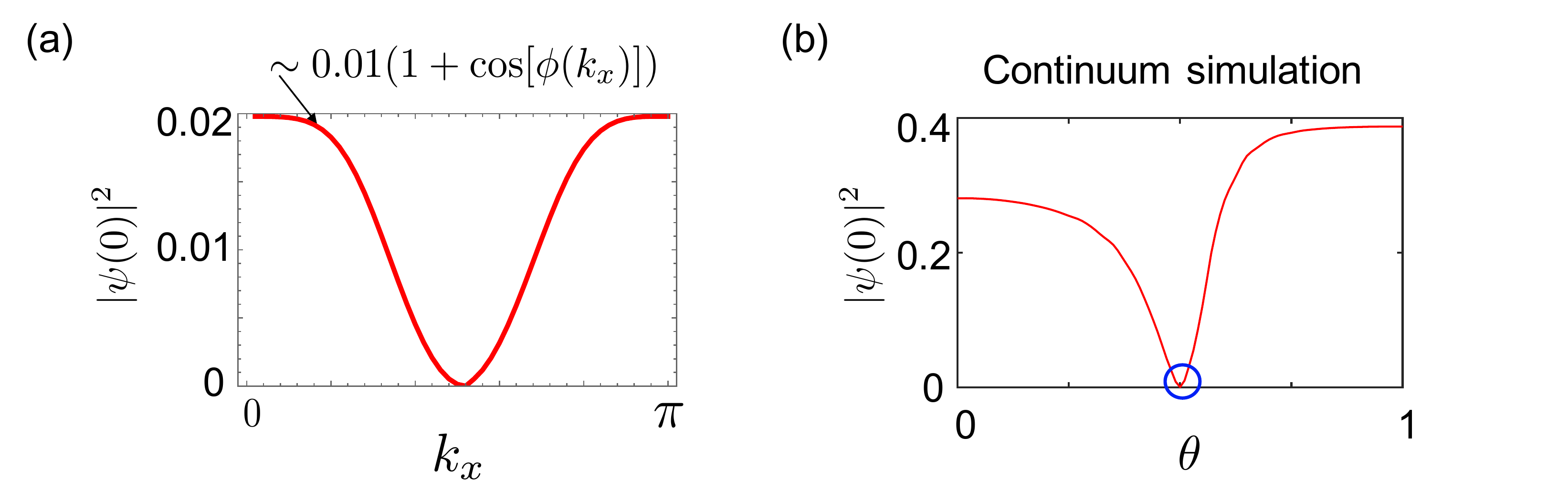}
\caption{(a) A tight-binding calculation of the intensity of the states with energy closest to zero at $y=0$ for $k_{x}\in[0,\pi]$. (b) A corresponding full-continuum waveguide simulation, where $\theta$ plays the role of $k_{x}$ and parameterizes the size of, and the distance between, waveguides. The blue circle identifies that there is a point where $|\psi(0)|^2=0$ and thus where $\phi=\pi$. }
\label{fig:expsim}
\end{figure}

More specifically, we now study the intensity for a gapless chain attached to the family of 1D insulating chains that correspond to  $k_{x}\in[0,\pi]$. To extract the information about $\phi$ from Eq.~\eqref{eq:density}, in Fig.~\ref{fig:expsim} (a) we plot the intensity at $y=0$ for $k_{x}\in[0,\pi],$ which should be proportional to $1+\cos\phi$ (recall $k=\pi/2$). We note that more details about the intensity over the whole gapless chain can be found in Appendix~\ref{sec:exp}. We find that the plot in Fig.~\ref{fig:expsim} (a) is approximately captured by $0.01(1+\cos\phi)$ from which we can observe that $\phi$ indeed winds by $2\pi$ when $k_{x}$ changes from $0$ to $\pi$. A corresponding full-continuum simulation for coupled waveguides is shown in Fig.~\ref{fig:expsim} (b), which qualitatively shows the same results. Interestingly, even though the reflection phase in the full-continuum simulation does not precisely match the tight-binding results because it does not exactly wind by $2\pi$ when $\theta$ changes from $0$ to $1$\footnote{This is ascribed to the subtlety of the connection between the gapless chain and the insulating chains discussed in Appendix~\ref{sec:calculation2} and \ref{sec:exp}.}, one can still see that $\phi$ crosses $\pi$ once (i.e., $|\psi(0)|^2$ touches zero as indicated by the blue circle) when $k_{x}$ changes from $0$ to $\pi$, as expected from the analysis in Sec~\ref{sec: relectionRTP}.  

In an experimental context the eigenstate closest to mid-gap, and the corresponding reflection phase, can be accessed and/or observed by placing an auxiliary waveguide sufficiently far away from the system \cite{noh2018topological}.
The auxiliary waveguide should be identical to the waveguides in the gapless region. Such a waveguide can act as an external drive to inject light into the main system without significant perturbations of the intrinsic eigenstates of the main system. This  results in the selective excitation of the eigenstates near $E=0$, for which we can immediately apply our theoretical analysis above.

\section{Conclusions and remarks}
\label{sec:conclusion}
In conclusion, we developed a scattering theory to understand delicate TIs.  Furthermore, we proposed a mechanism to stabilize surface states of delicate TIs on non-sharp boundaries. 
To test our predictions we proposed a photonic experiment to measure the reflection phase of delicate TIs. Since a nonzero RTP indicates an intrinsically polarized system that breaks inversion symmetry, such materials potentially have large optical effects like the photovoltaic effect discussed recently in Ref. \onlinecite{alex:2022topological}. Thus, it is important and interesting to search for such kinds of materials in solid state and metamaterial platforms. The scattering theory discussed in this work can be used as an important indicator for the search. We hope that in the future, the experimental implementation of delicate TIs and the measurement of their reflection phase can be done not only in photonic systems, but also in acoustic systems, electric circuits, and even quantum materials.

\section*{Acknowledgements}
P.~Z. thanks Yidong Chen, Tomáš Bzdušek, Aris Alexandradinata, Sachin Vaidya, and Xiaoqi Sun for insightful discussions. P.~Z. and T.~L.~H. thank the US Office of Naval Research (ONR) Multidisciplinary University Research Initiative (MURI) grant N00014-20- 1-2325 on Robust Photonic Materials with High-Order Topological Protection
for support.
\appendix

\section{Calculation of the reflection phase in a 1D chain}
\label{sec:reflectioncalculation}

The setup of the scattering problem on a 1D chain is shown in Fig.~\ref{fig:scattering} (a). We label the positions of the lattice sites of the target 1D system in the range $[1,N]$, and connect its two boundaries to reservoirs. For simplicity, we take the reservoir to be a simple 1D gapless chain with Hamiltonian $H_{\text{gapless}}=\sum_{n}\left[(h_{0}c_{n+1}^{\dag}c_{n}+h.c.)-\mu c_{n}^{\dag}c_{n}\right].$ The spectrum is given by $E=-\mu+2 h_{0}\cos(k)$, where $\mu$ is the chemical potential of the whole system, i.e., the target system and the reservoirs have the same chemical potential $\mu$. For simplicity, we choose the lattice constant to be $1$ and $\mu=0$. 

To calculate the scattering matrix $S$ in Eq.~\eqref{eq:S}, we define a transfer matrix $T$ such that
\begin{equation}
\label{eq:T}
    T\begin{pmatrix}
    B
    \\
    A
    \end{pmatrix}=\begin{pmatrix}
    T_{11} & T_{12}
    \\
    T_{21} & T_{22}
    \end{pmatrix}\begin{pmatrix}
    B
    \\
    A
    \end{pmatrix}=\begin{pmatrix}
    D
    \\
    C
    \end{pmatrix}.
\end{equation}
Then,
\begin{equation}
\label{eq:reflection}
\begin{aligned}
&r=-T_{11}^{-1}T_{12}, \ t=-T_{21}T_{11}^{-1}T_{12}+T_{22},
\\
&r^{\prime}=T_{21}T_{11}^{-1}, \ t^{\prime}=T_{11}^{-1}.
\end{aligned}
\end{equation}

For the central region we consider a general short-range 1D tight-binding model with Hamiltonian:
\begin{equation}
\label{eq:1Dhamil}
H(s)=\sum_{n=1}^{N-1}( c^{\dag}_{n+1} h_{n+1,n}(s) c_{n}+\text{h.c.})+h_{n}^{\prime}(s)c^{\dag}_{n}c_{n},
\end{equation}
where $h_{n+1,n}$ and $h^{\prime}_{n}$ are generically matrices, and $s$ is a periodic parameter that will be used when we consider 2D systems as pumps of a 1D chain.

To calculate the transfer matrix, we look at the static Schr$\ddot{\rm{o}}$dinger equation given by $H(s)$:
\begin{equation}
\label{eq:sequation}
H(s)\sum_{n}\psi_{n}\ket{n}=E\sum_{n}\psi_{n}\ket{n},
\end{equation}
which leads to 
\begin{equation}
\label{eq:sequation1}
0=h_{n,n-1}(s)\psi_{n-1}+h^{\dag}_{n+1,n}(s)\psi_{n+1}+(h_{n}^{\prime}(s)-E)\psi_{n}.
\end{equation}
Note that we focus on the scattering process for input waves at the Fermi level, i.e., $E=\mu=0$. Then, the above equation can be rewritten as
\begin{equation}
\label{eq: transfer}
\begin{pmatrix}
\psi_{n}
\\
\psi_{n+1}
\end{pmatrix}=\mathcal{M}_{n}\begin{pmatrix}
\psi_{n-1}
\\
\psi_{n}
\end{pmatrix},
\end{equation}
where
\begin{equation}
\label{eq: M}
\mathcal{M}_{n}=\begin{pmatrix}
0 & 1
\\
-(h^{\dag}_{n+1,n}(s))^{-1}h_{n,n-1}(s) & -(h^{\dag}_{n+1,n}(s))^{-1}h^{\prime}_{n}(s)
\end{pmatrix}.
\end{equation}
Note that we assume the $h_{n+1,n}$ to be invertible. For non-invertible $h_{n+1,n}$, one needs use the singular-value-decomposition (SVD) method as discussed in Ref. \onlinecite{dwivedi2016bulk}. At the interfaces between the target system and the reservoirs, we apply continuity of the eigenstates to find 
\begin{equation}
\label{eq:boundary1}
\begin{pmatrix}
\psi_{0}
\\
\psi_{1}
\end{pmatrix}=\Lambda\theta_{L}\begin{pmatrix}
B
\\
A\end{pmatrix}, \ \begin{pmatrix}
\psi_{N}
\\
\psi_{N+1}
\end{pmatrix}=\Lambda\theta^{-1}_{R}\begin{pmatrix}
D
\\
C
\end{pmatrix},
\end{equation}
where
\begin{equation}
\label{eq:L}
\Lambda=\begin{pmatrix}
\mathbbm{1} & \mathbbm{1}
\\
e^{-ik}\mathbbm{1} & e^{ik}\mathbbm{1} \end{pmatrix},
\end{equation}
and 
\begin{equation}
\label{eq:theta}
\theta_{L}=\begin{pmatrix}
e^{ikx_{L}}\mathbbm{1} & 0
\\
0 & e^{-ikx_{L}}\mathbbm{1}
\end{pmatrix}, \ \theta_{R}=\begin{pmatrix}
e^{ik(N-x_{R})}\mathbbm{1} & 0
\\
0 & e^{-ik(N-x_{R})}\mathbbm{1}
\end{pmatrix}.
\end{equation}
Comparing Eq.~\eqref{eq:T}, \eqref{eq: transfer}, and \eqref{eq:boundary1}, we can write the transfer matrix $T$ as 
\begin{equation}
\label{eq:T1}
T=\theta_{R} \Lambda^{-1}\left(\prod_{n=1}^{N}\mathcal{M}_{n}\right)\Lambda\theta_{L}.
\end{equation}
With $T$ calculated in Eq.~\eqref{eq:T1}, we apply Eq.~\eqref{eq:reflection} to get the reflection coefficient and subsequently its phase.

In our calculation, input states with $E=0$ in the gapless chain with dispersion $E=2 h_{0}\cos k$ have momentum $k=\pi/2$. At the interface between gapless chains and the target system, we have $h_{1,0}=h_{0}$ and $h_{N+1,N}=h_{0}$. Since in Eq.~\eqref{eq:T1}, $\mathcal{M}_{n}$ matrices connect site $0$ and site $N+1$, it is reasonable to regard $0$ and $N+1$ as the reflection point and thus $x_{L}=0$ and $x_{R}=N+1$. This corresponds to 
\begin{equation}
\label{eq:theta1}
\theta_{L}=\begin{pmatrix}
\mathbbm{1} & 0
\\
0 & \mathbbm{1}
\end{pmatrix}, \ \theta_{R}=\begin{pmatrix}
-i\mathbbm{1} & 0
\\
0 & i\mathbbm{1}
\end{pmatrix}.
\end{equation}
\subsection{Reflection matrix of the MRW model in Eq.~\eqref{eq: MRW}}\label{subsec:reflectioncalculationMRW}
For the 3D model in Eq.~\eqref{eq: MRW}, we can read off the real space hopping (in the $z$-direction) after a Fourier transformation of the Bloch Hamiltonian. This model has just nearest-neighbor tunneling in the $z$ direction. For labeling purposes, in this subsection $n$ refers to the unit cell label, with implicit labeling of sites within a unit cell. As a result, the hopping amplitudes are matrices, and they are the same for even and odd unit cells within the central region, namely $h_{n+1,n}=h_{n,n-1}$. Using this convention we have the matrix elements of $ h_{n,n+1}$ as
\begin{equation}
\begin{aligned}
\left(h_{n,n+1}\right)_{11}=& u - \cos(k_x) - \cos(k_y), \\
\left(h_{n,n+1}\right)_{12}=& 2 \imath \sin(k_x) - 2 \sin(k_y), \\
\left(h_{n,n+1}\right)_{21}=&0, \\
\left(h_{n,n+1}\right)_{22}=&-u + \cos(k_x) + \cos(k_y).
\end{aligned}
\end{equation}
The matrix elements of $ h'_{n}$ are
\begin{equation}
\begin{aligned}
\left(h'_n \right)_{11}=& 2 \cos (k_x) (u-\cos (k_y))-\cos (2 k_x)\\
&+2 u \cos (k_y)-\cos (2 k_y)-u^2-1, \\
\left(h'_n \right)_{12}=&2 (-\sin (k_y)+\imath \sin (k_x)) (\cos (k_x)+\cos (k_y)-u), \\
\left(h'_n \right)_{21}=&2 (\sin (k_y)+\imath \sin (k_x)) (-\cos (k_x)-\cos (k_y)+u), \\
\left(h'_n \right)_{22}=&2 \cos (k_x) (\cos (k_y)-u) \\
&+\cos (2 k_x)-2 u \cos (k_y)+\cos (2 k_y)+u^2+1.
\end{aligned}
\end{equation}
Notice that the above two matrices are both invertible. We can use Eq.~\eqref{eq: transfer} to calculate the matrix $\mathcal{M}_n$, treating the hopping strength $h_{n,n+1}$, $h'_n$ as matrices. The rest follows.

\subsection{Reflection matrix of the 2D model in Eq.~\eqref{eq:2DRTP}}
\label{sec:calculation2}

For the 2D model in Eq.~\eqref{eq:2DRTP} the hoppings and onsite potentials are illustrated in Fig.~\ref{fig:2DRTP}(a). If we use the same method as in Appendix~\ref{subsec:reflectioncalculationMRW}, we find the hopping matrices are singular. We can either use the singular value decomposition method to calculate the canonical $n$-channel transfer matrix (where $n$ is the number of internal degrees of freedom in a unit cell) as is discussed in Ref. \onlinecite{dwivedi2016bulk}, or we can make a simplification assuming the incoming wave is a single channel. Since our model is an SSH-type 1D chain, the transfer matrix of single channel incoming waves has the benefit of simplicity.

Therefore, in this subsection, we will consider single channel incoming waves, where we have $h_{n+1,n}(s=k_{x})$ equal to $m\sin k_{x}$ ($\sin k_{x}$) for odd (even) $n$, and  $h_{n}^{\prime}=(-1)^{n-1}\cos k_{x}$. Since we are considering only single channel incoming waves, here $n$ label sites instead of unit cells, and $h$'s are now complex numbers instead of matrices. In doing so we lose information about the unit cell structure so we will need to enforce the total number of sites to be an even number by hand so that we have translation symmetry. Assuming the hoppings are all non-zero, in the target system we can write the matrices 
\begin{equation}
\label{eq:2D M mat}
\mathcal{M}_{2n+1} =\begin{pmatrix}
0 & 1
\\
-\frac{1}{m} & -\frac{\cos(k_x)}{m\sin(k_x)}
\end{pmatrix},
\mathcal{M}_{2n} =\begin{pmatrix}
0 & 1
\\
-m & \frac{\cos(k_x)}{\sin(k_x)}
\end{pmatrix}.
\end{equation}

The simplicity of these matrices hides a subtlety when it comes to choosing the form of the hopping strength between the reservoir and the central region. Notice the transfer matrix here has a natural singularity whenever the hopping vanishes, i.e., when $\sin(k_x)=0$. This corresponds to a fully decoupled case, where the intermediate matrices will contain zero matrices. Our transfer matrix method will no longer work in this completely decoupled case. However, we may consider the limit where $\sin(k_x)\to0$. In this limit the matrices in Eq.~\eqref{eq:2D M mat} both take the form $\begin{pmatrix}
0&0\\0&1
\end{pmatrix}$ up to an overall factor. This matrix form is maintained when taking products with itself regardless of the number of sites in the central region.

\begin{figure}[h]
\centering
\includegraphics[width=0.6\columnwidth]{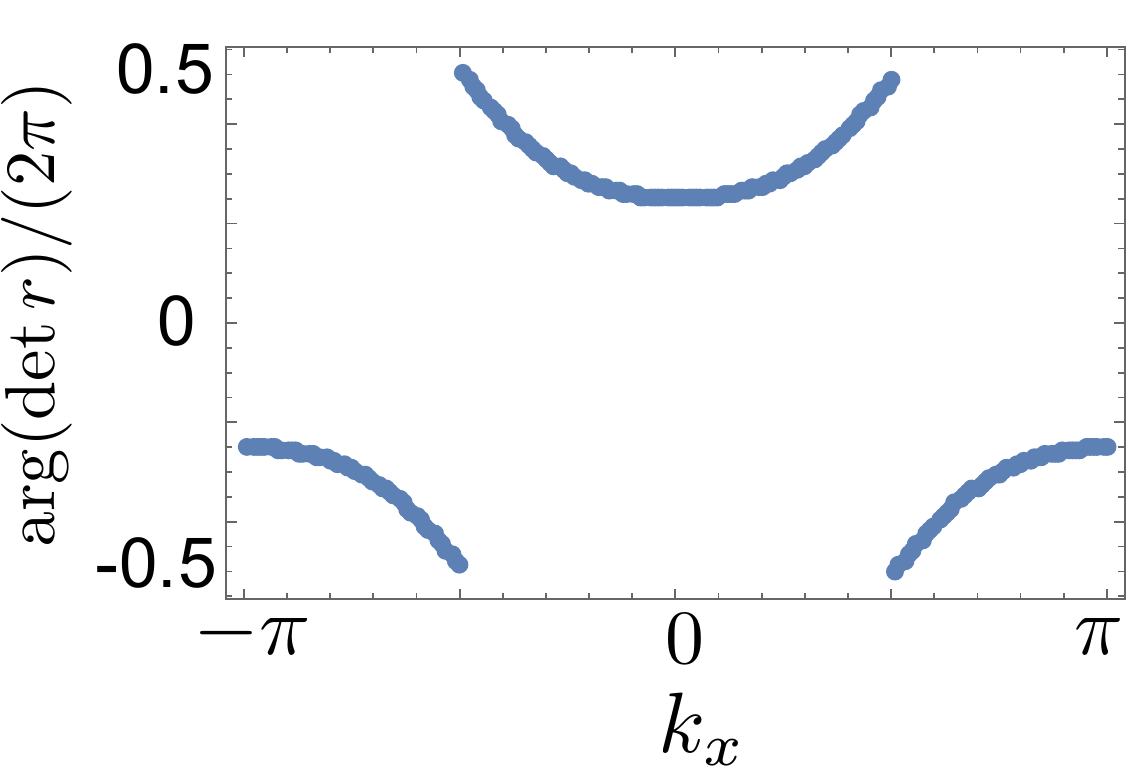}
\caption{The reflection phase for the model in Eq.~\eqref{eq:2DRTP} when connecting to the metallic chain with a hopping comparable to the onsite potential.}
\label{fig:reflectiondifferentconnect}
\end{figure}

When connected to the reservoir, we have a choice to make. If we set the gapless reservoir chain's hopping strength to be a constant that is comparable with the strength of the onsite potential at $k_{x}=0$ or $\pi$, then at the interface when $\sin(k_x)\to0$ the matrix $\mathcal M$ is of the form $\begin{pmatrix}
0&0\\ \pm1&1
\end{pmatrix}$ up to an overall factor. This leads to a resultant reflection phase of $\pm\pi/2$. We numerically verify the predicted number of crossings with $\pi$ of the reflection matrix eigenvalues as shown in Fig. \ref{fig:reflectiondifferentconnect}:  When we go from $k_{x}=0$ to $k_{x}=\pi$, the reflection phase varies from $+\pi/2$ to $-\pi/2$ continuously, and crosses $\pi$ once. Still, there is a small discrepancy from the conclusion we drew from the multi-channel case, which is that the values of the phase at $k_{x}=0,\pi$ are not integer multiples of $2\pi$ as predicted in the multichannel case, they are $\pm \pi/2$. However, if we go back to the two-channel case, where we have one more channel connected to another sublattice in the boundary unit cell, then at $k_{x}=0$ and $k_{x}=\pi$, the reflection matrices are just $\begin{pmatrix}
e^{-i\pi/2} & 0
\\
0 & e^{i\pi/2}
\end{pmatrix}$ and $\begin{pmatrix}
e^{i\pi/2} & 0
\\
0 & e^{-i\pi/2}
\end{pmatrix}$, and the determinants are $1$ which is consistent with the argument in the main text. As a consequence, if we consider only single-channel scattering and set the hopping to a constant comparable with the strength of onsite potential at $k_{x}=0,\pi$, we cannot see the returning winding number of the phase.

To fix this, we will prescribe a better way of setting the reservoir coupling that
is free of the aforementioned discrepancy.  
To accomplish this we can vary the hopping strength between the gapless chain and the insulating chain in accordance with the change of hopping strength in the insulating chain, i.e., we vary the gapless chain's hopping strength as $\sin^2(k_x)$. This smooth choice of the reservoir's hopping strength goes to zero simultaneously with the insulating chain. At the interface when $\sin(k_x)\to0$ the matrix $\mathcal M$ now takes the form $\begin{pmatrix}
0&0\\0&1
\end{pmatrix}$ up to an overall factor. This will lead to a phase which is an integer multiple of $2\pi$. In this case, we can nicely observe the returning winding even in a single-channel setup. 

The deviation from returning winding at $k_{x}=0,\pi$ can also be removed by setting the hopping between the gapless chain and the insulating chain to a small value compared to the onsite potential achievable in the insulating chain. Doing so ensures the $\mathcal M$ matrices approximately retain the form in the decoupled limit in the insulating chain, which again leads to the observation of the returning winding as shown in the main text.  In realistic cases, the hopping between the gapless chain and the insulating chain can be smaller than the onsite potential achievable in the insulating chain but might not be much smaller. In this case  we would expect to observe a reflection phase between $0$ and $\pi/2$ at $k_{x}=0,\pi$, and an approximate returning winding.

\section{Local potential on the boundary}
\label{sec:boundarypotential}
Given the Hamiltonian in Appendix~\ref{sec:calculation2}, we add boundary potential $V=v_{1} c^{\dag}_{1}c_{1}+v_{N}c^{\dag}_{N}c_{N}$. For the plots in Fig.~\ref{fig:gboundary}, we use $v_{1}=-v_{N}=1.2$ which breaks the chiral and mirror symmetries at both $M$ and $\Gamma$. For the plots in Fig.~\ref{fig:mboundary}, we use $v_{1}=v_{N}$ which breaks the chiral symmetry but preserves the mirror symmetry. In Fig.~\ref{fig:mboundary} (a), (b), and (c), we use $v_{1}=v_{N}=1.2,-1.2, 0.5$, respectively.

\begin{widetext}
\section{Supplementary discussion about the experimental proposal}
\label{sec:exp}

\begin{figure}[t]
\centering
\includegraphics[width=1\columnwidth]{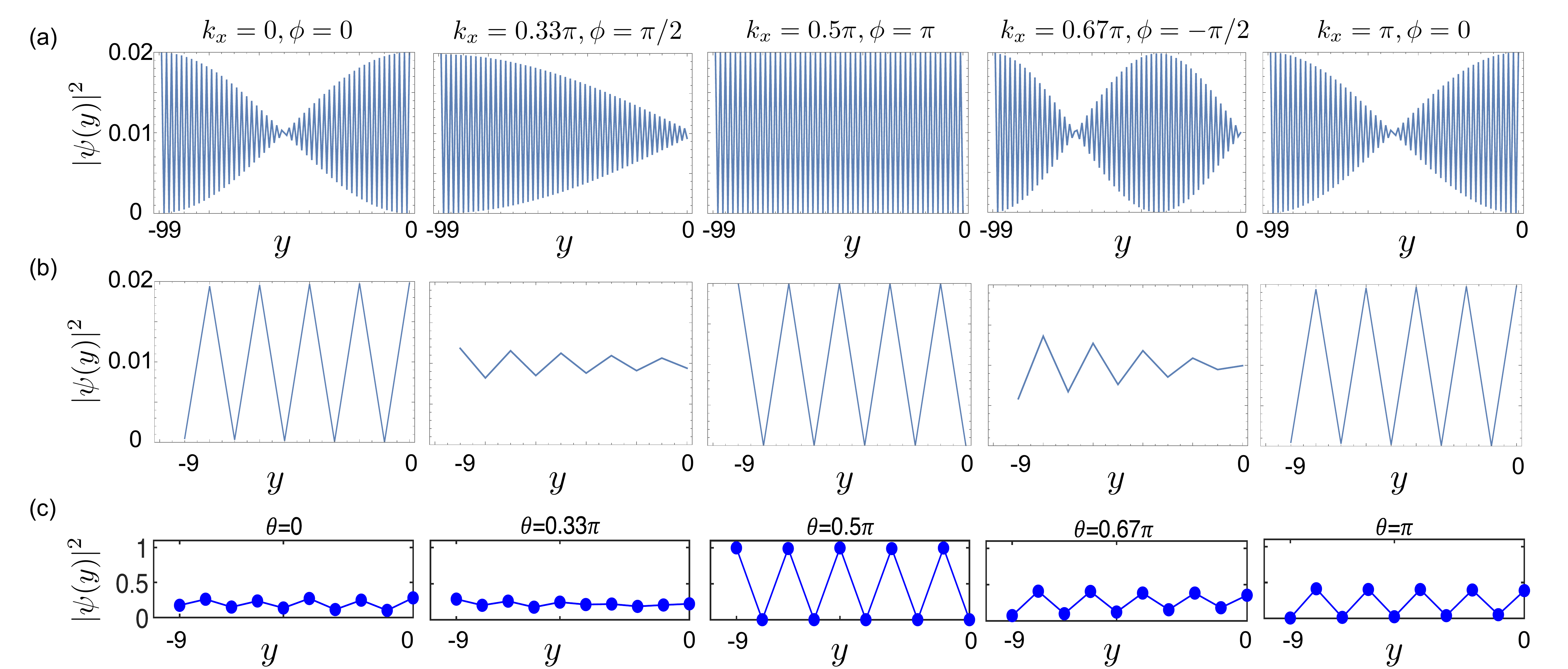}
\caption{(a) shows the intensity of the states with energy closest to zero in the 100 site gapless chain attached to insulating chains corresponding to different $k_{x}\in[0,\pi]$. (b) shows the density for the ten sites closest to the insulating chain. (c) shows a full-continuum waveguide simulation, where $\theta$ parametrizes the size of and the distance between waveguides and plays the role of $k_{x}$.}
\label{fig:expsim1}
\end{figure}
Here, we show that the intensity over the gapless chain is indeed captured by Eq.\eqref{eq:density} for our setup in the main text. Since both the gapless chain and the insulating chain are finite (100 and 10 unit cells respectively), we have eigenstates close to $E=0,$ but not exactly at $E=0.$ Additionally, the associated  wave vector is not exactly $k=\pi/2,$ but instead is $k=\pi/2+\delta k$ with $\delta k\sim 2\pi/100 \ll \pi/2$. Substituting $k=\pi/2+\delta k$ into Eq.~\eqref{eq:density}, we can derive the intensity to be $|\psi(y)|^2\sim 1+\cos(\pi y+2\delta kn-\phi)=1+\cos(2\delta k y-\phi)\cos\pi y$ up to a normalization factor. 
As shown in Fig.~\ref{fig:expsim1} (a), our tight-binding calculations of intensity for eigenstates of the gapless chain closest to $E=0$ are consistent with this result. Indeed at the five typical $k_{x}$ values between $0$ and $\pi$ that we show, the intensity plots are essentially the plots of $0.01(1+\cos((\pi+2\delta k)y-\phi)\cos\pi y)$ with $(\delta k=\pi/100,\phi=0)$, $(\delta k=\pi/200,\phi=\pi/2)$, $(\delta k=0,\phi=\pi)$, $(\delta k=3\pi/200,\phi=-\pi/2)$, and $(\delta k=\pi/100,\phi=0).$ Crucially, the series of intensities indicate a $2\pi$ winding of $\phi$ as the effective $k_x$ goes from $0$ to $\pi$. 

We show the results of the reverting process from $k_{x}=\pi$ to $k_{x}=2\pi$ in Fig.~\ref{fig:expsimreturning}, which manifest a $-2\pi$ winding (returning winding) of the reflection phase $\phi$ from $k_{x}=\pi$ to $k_{x}=2\pi$. 
We focus on $y=0$ in the main text, but we can also extract information about $\phi$ by looking at a part of the gapless chain with a length much shorter than the total length of the gapless chain. If we choose say the ten sites closest to the insulating chains (i.e., sites with coordinate labels $y\in [-9,0]$) in the gapless chain, then the variation due to $\delta k$ can be ignored. Then, $\phi$ can be derived by simply comparing our plots with the function $1+\cos (\pi y -\phi)$ because it has the largest oscillation when $\phi=0,\pi$, and gradually becomes flat when $\phi$ goes to $\pm \pi/2$, as shown in Fig.~\ref{fig:expsim} (b). A corresponding full-continuum simulation for coupled waveguides is shown in Fig.~\ref{fig:expsim} (c), which qualitatively shows the same results.

\begin{figure}[h]
\centering
\includegraphics[width=1\columnwidth]{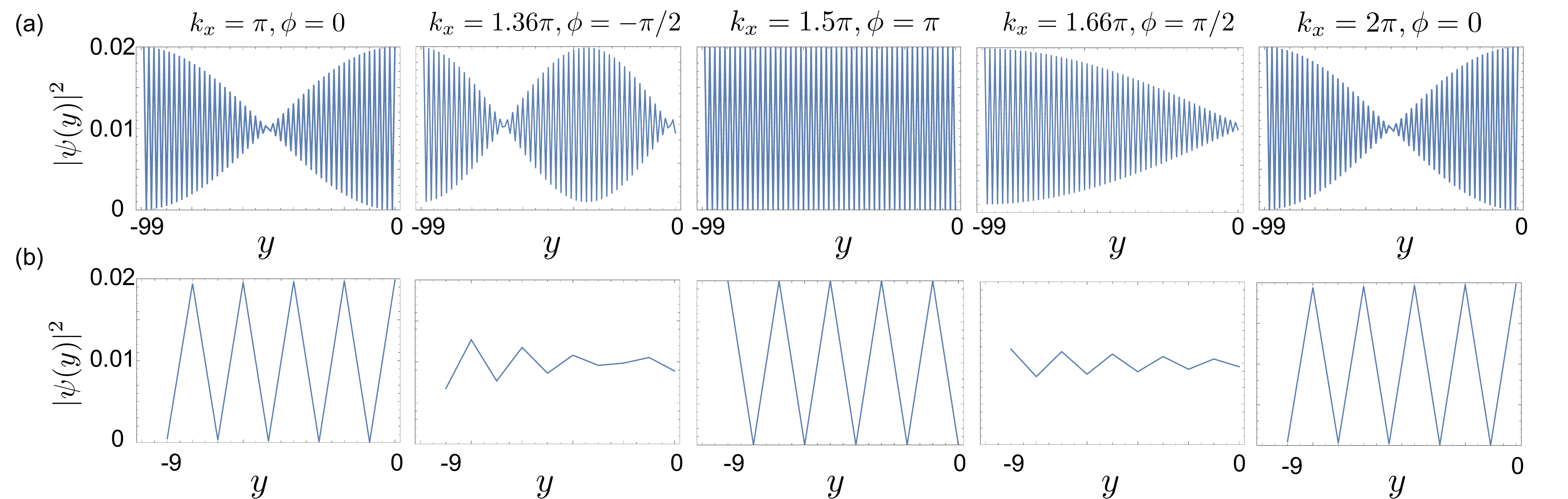}
\caption{(a) shows the intensity of the states with energy closest to zero in the 100 site gapless chain attached to insulating chains corresponding to different $k_{x}\in [\pi,2\pi]$. (b) shows the density for the ten sites closest to the insulating chain.}
\label{fig:expsimreturning}
\end{figure}

Next, we discuss the subtlety of the distance between the gapless chain and different insulating chains, which is related to the subtlety discussed in the tight-binding calculation discussed in Appendix~\ref{sec:reflectioncalculation}, i.e., the  reflection phase we observe depends on how we connect the gapless chain and the insulating chain. In the full-continuum simulation, we fix the distance between the gapless chain and the insulating chain to be $d_{0}$, which is the distance between adjacent waveguides in the gapless chain. Thus, the coupling between the gapless chain and the insulating chain is the same as the coupling between adjacent sites in the gapless chain. As shown in Fig.~\ref{fig:coupling}, the onsite potential oscillation amplitude (i.e., the onsite potential proportional to $\sigma_{z}$ in the tight-binding Hamiltonian) is twice the coupling between the gapless chain and the insulating chain (i.e., not that much larger than the coupling between the gapless chain and the insulating chain). Thus, as discussed in Appendix~\ref{sec:calculation2}, from $\theta=0$ to $\theta=\pi$, the reflection phase does not exactly wind by $2\pi$, which is consistent with what we show in Fig .~\ref{fig:expsim} (b) and \ref{fig:expsim1} (c).
\begin{figure}[h]
\centering
\includegraphics[width=0.7\columnwidth]{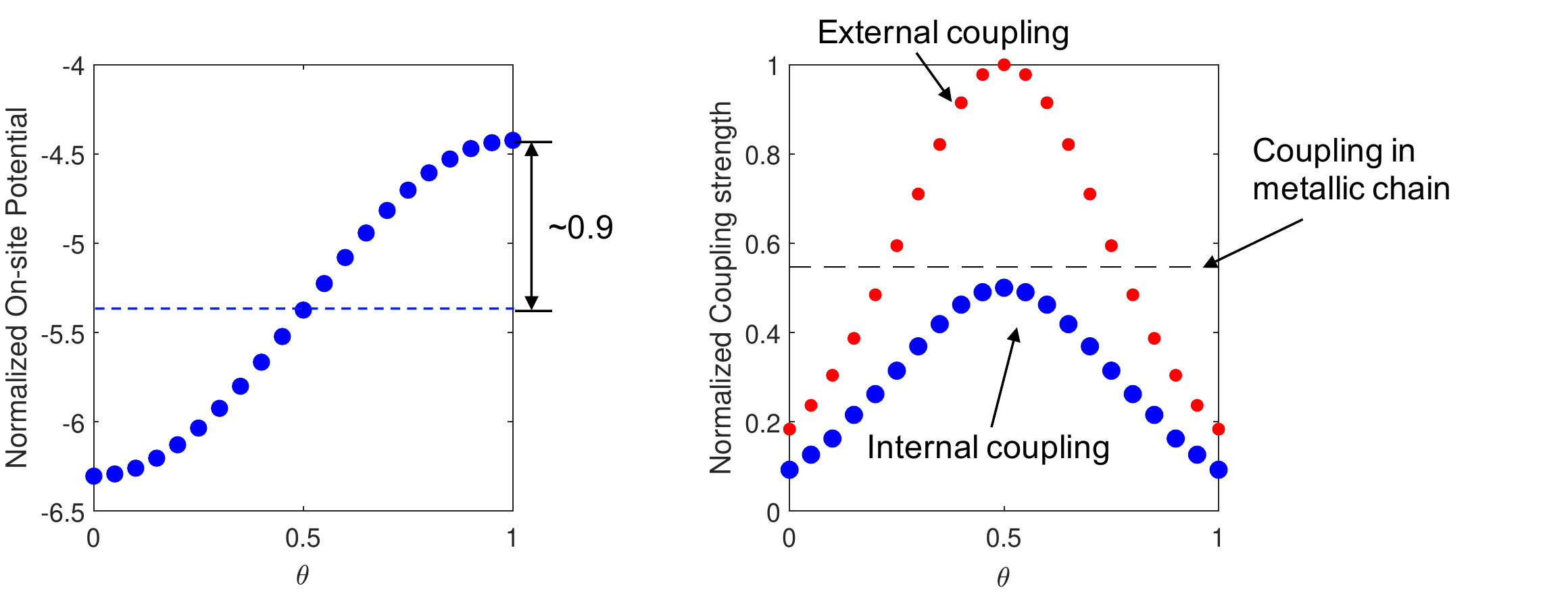}
\caption{The left panel shows the on-site potential with oscillation amplitude $\sim 0.9$. The right panel shows the internal and external couplings in the insulating chain and the couplings in the gapless chain. Note that all energy scales are normalized with respect to the external coupling strength in the insulating chain at $\theta=0.5$.}
\label{fig:coupling}
\end{figure}
\end{widetext}

\bibliography{apssamp}




\end{document}